# Theoretical realization of two-dimensional $M_3(C_6X_6)_2$(M= Co, Cr, Cu, Fe, Mn, Ni, Pd, Rh and X= O, S, Se) metal-organic frameworks

Bohayra Mortazavi[*b,c], Masoud Shahrokhi[d], Tanveer Hussain[e,f], Xiaoying Zhuang[c] and Timon Rabczuk[a]

[a] *Institute of Research & Development, Duy Tan University, Quang Trung, Da Nang, Vietnam.*

[b] *Institute of Structural Mechanics, Bauhaus-Universität Weimar, Marienstr. 15, D-99423 Weimar, Germany.*

[c] *Cluster of Excellence PhoenixD (Photonics, Optics, and Engineering–Innovation Across Disciplines), Gottfried Wilhelm Leibniz Universität Hannover, Hannover, Germany.*

[d] *Department of Physics, Faculty of Science, Razi University, Kermanshah, Iran.*

[e] *School of Molecular Sciences, The University of Western Australia, Perth, WA 6009, Australia. Centre for*

[f] *Theoretical and Computational Molecular Science, Australian Institute for Bioengineering and Nanotechnology, The University of Queensland, Brisbane, Queensland 4072, Australia*

**Abstract**

Two-dimensional (2D) conductive metal–organic framework (MOF) lattices have recently gained remarkable attentions because of their outstanding application prospects. Most recently, Cu-hexahydroxybenzene MOF was for the time experimentally realized, through a kinetically controlled approach. Cu-HHB belongs to the family of conductive MOFs with a chemical formula of $M_3(C_6X_6)_2$(X=NH, O, S). Motivated by the recent experimental advance in the fabrication of Cu-HHB, we conducted extensive first-principles simulations to explore the thermal stability, mechanical properties and electronic characteristics of $M_3(C_6X_6)_2$(M= Co, Cr, Cu, Fe, Mn, Ni, Pd, Rh and X= O, S, Se) monolayers. First-principles results confirm that all considered 2D porous lattices are thermally stable at high temperatures over 1500 K. It was moreover found that these novel 2D structures can exhibit linear elasticity with considerable tensile strengths, revealing their suitability for practical applications in nanodevices. Depending on the metal and chalcogen atoms in $M_3(C_6X_6)_2$ monolayers, they can yield various electronic and magnetic properties, such as; magnetic semiconducting, perfect half metallic, magnetic and nonmagnetic metallic behaviours. This work highlights the outstanding physics of $M_3(C_6X_6)_2$ 2D porous lattices and will hopefully help to expand this conductive MOF family, as promising candidates to design advanced energy storage/conversion, electronics and spintronics systems.

Corresponding authors: *bohayra.mortazavi@gmail.com, #timon.rabczuk@uni-weimar.de;

## 1. Introduction

Among the various classes of materials, two-dimensional (2D) nanomaterials are currently among the most interesting materials owing to their wide-range application prospects, like post-silicon nanoelectronics, nanooptics, bio- and nano-sensors, thermal management devises, next generation energy storage/conversion systems and structural components in aerospace. The significance of 2D materials came into consideration only after the great success of graphene [1,2], which exhibits uniquely high tensile strength [3] and thermal conductivity [4,5] along with exceptional and tuneable optical and electronic characteristics [6–10]. Since then, different classes of 2D materials were born and have been continuously expanding. Despite of very attractive properties of graphene, it has some limitations for various applications. These limitations of graphene have been acting as great motivations to astonishingly expand the 2D materials family during the last decade. For example, graphene does not show an electronic band-gap in its pristine form, which have promoted the design and fabrication of novel 2D materials with inherent semiconducting electronic character, like: transition metal dichalcogenides such as the $MoS_2$ and $WS_2$ [11–15], phosphorene [16,17] and 2D carbon nitrides [18–24]. As another example, densely packed atomic lattice of graphene with all $sp^2$ carbon atoms, also decreases its capacity for the storages of various metal-ions, which is a drawback for energy storage systems, like Li-ion batteries. In this case, other carbon based 2D materials with porous atomic structures, like graphdiyne [25] lattices with hybrid sp and $sp^2$ carbon atoms have been recently synthesized [26–28], which exhibit superior storage capacities than graphene for the application in various rechargeable metal-ion battery technologies [29,30].

Conductive metal-organic frameworks (MOF) are among the most appealing classes of 2D materials, which show porous atomic lattices with intrinsic electrical conductivity and integrated functionalities. In these structures, longrange electron delocalization occurs throughout the framework because of the conjugation of the organic linker and forming strong orbital interaction with the central connecting metal atoms. The porous structures of conductive MOFs also provide efficient conditions for the access to active sites in the materials, which can be ideal for adatom adsorptions and diffusion in energy storage systems. Previous experimental studies have confirmed that conductive MOFs can serve as promising candidates in many applications, including electrocatalysis [31,32], energy storage systems [33,34], sensors [35,36] and electronic devices [37,38]. Nevertheless, synthesize of high quality and large-area 2D MOFs are yet among the most critical challenges toward their practical applications. In response to these challenges, in a latest experimental advance, Feng *et al.* [33] succeeded in the fabrication of hexaaminobenzene (HAB)-derived nanomembranes with chemical formulas of $M_3$-$C_{12}(NH)_{12}$(M=Cu, Ni). According to their experimental tests, Cu- and Ni-HAB nanosheets can yield highly desirable performances as supercapacitors, because of their excellent chemical stability in acidic and aqueous solutions, high volumetric and areal capacitances, remarkable reversible redox behaviours and acceptable cycling stability [33]. In another recent experimental achievement, Park *et al.* [39] synthesized Co-HAB nanosheets with excellent performances for the application as an electrode material for sodium-ion storage. Park *et al.* [40] also most recently fabricated Cu-hexahydroxybenzene (Cu-HHB) a novel conductive MOF in the 2D form, through a kinetically controlled approach with a competing coordination reagent. These recent and continuous experimental advances [33,39,40] with respect to the synthesize of conductive MOFs nanosheets, with the chemical formula of $M_3(C_6X_6)_2$(X=NH, O), reveal substantial steps toward their practical applications in various advanced and critical technologies. Successful fabrication of these novel nanomaterials consequently highlights the importance of new studies to provide in-depth understanding of their intrinsic properties. Moreover, these accomplishments also raise a simple question concerning the stability, intrinsic properties and possibility of the synthesis of other conductive 2D MOFs with similar atomic lattices, but with different metal and chalcogen atoms. Because of the complexities of experimental fabrication and characterization techniques, theoretical studies can be considered as the fastest approaches to examine new compositions and estimate their properties and suggest possible synthesis routes [41–43]. In this short communication, our objective is to investigate thermal stability, mechanical properties and electronic characteristics of $M_3(C_6X_6)_2$(M= Co, Cr, Cu, Fe, Mn, Ni, Pd, Rh and X= O, S, Se) monolayers using the first-principles density functional theory simulations. Worthy to note that this study is a complementary to our previous investigation [44], regarding the similar MOF lattices, $M_3(C_6X_6)_2$(X=NH). The acquired results by our first-principles modelling highlight attractive properties of this novel class of 2D materials and will hopefully lead to practical application of these novel nanomaterials in the next generation electronic and energy storage systems [10,45].

## 2. Computational methods

Spin polarized density functional theory (DFT) calculations have been performed employing the *Vienna Ab-initio Simulation Package* (VASP, version 5.4.1) [46–48]. The exchange-correlation effects were treated within generalized gradient approximation within the Perdew-Burke-Ernzerhof (PBE) functional [49]. Projector-augmented wave (PAW) [50] potentials as provided by the VASP have been used for all elements. We used a plane-wave cutoff energy of 500 eV and the convergence criterion for the electronic self consistence-loop was set to be $10^{-4}$ eV. VESTA [51] package was used for the illustration of atomic structures and charge densities as well. Periodic boundary conditions were applied along the all three Cartesian directions, with a vacuum layer of 16 Å to avoid image-image interactions along the monolayers thickness. Energy minimized structures were acquired by altering the size of the unit-cells and then employing the conjugate gradient method for the geometry optimizations using a 3×3×1 Monkhorst-Pack [52] k-point mesh size. The convergence criterion for the Hellmann–Feynman forces on each atom was taken to be 0.01 eV/Å. In the evaluation of the electronic structure, a denser k-point mesh size of 11×11×1 was used. Worthy to note that since the considered structures include light or medium-sized atoms, the effects of spin-orbit coupling should be negligible [53], nonetheless this effect can be investigated in the future studies. Mechanical properties were evaluated by performing uniaxial tensile simulations [44]. Thermal stability of the MOF monolayers was examined by conducting the ab-initio molecular dynamics (AIMD) simulations for the energy minimized unit-cells, using the Langevin thermostat with a time step of 1 fs and 2×2×1 k-point mesh size [44].

## 3. Results and discussions

In Fig.1, samples of energy minimized $M_3(C_6X_6)_2$ monolayers with graphene-like hexagonal atomic lattices are shown. All studied nanosheets, exhibit similar atomic configurations, however, by increasing the mass of the chalcogen atoms the lattice constants increase, which is due to the elongation of the connecting bonds to the chalcogen atoms. It was found that the bonds in the hexagonal carbon rings remained convincingly unchanged for the all considered nanosheets. Interestingly, when comparing the $M_3(C_6O_6)_2$ monolayers with hexaaminobenzene $M_3(C_6N_6H_6)_2$ MOFs [44], very close atomic lattices can be found. To briefly analyse the bonding nature in these systems, we also depicted the electron localization function (ELF) [54] for the unit-cells in Fig. 1. ELF is a spatial function and takes a value between 0 and 1. As it is clear, high electron localization occur around the center of all C-C bonds, confirming the covalent bonding. For the all monolayers, noticeable electron localization happen around the chalcogen atoms, contrasting with completely delocalized electrons around the metal atoms. These reveal the existence of ionic bonding between the chalcogen and connecting metal atoms, which is more apparent in the case of $M_3(C_6O_6)_2$ nanosheets. As it can be seen in Fig. 1, in the cases of $M_3(C_6S_6)_2$ and $M_3(C_6Se_6)_2$ monolayers, between the chalcogen and metal atoms the electron localizations are more noticeable as compared with $M_3(C_6O_6)_2$ counterparts. This observation can be partially explained by around 25% lower electronegativity of S and Se atoms in comparison with O atoms and can reveal the increase in the contribution of covalent bonding in $M_3(C_6S_6)_2$ and $M_3(C_6Se_6)_2$ nanomembranes. In Table 1, the lattice constants of energy minimized structures and chalcogen-metal and chalcogen-carbon bond lengths in the considered monolayers are summarized. Worthy to note that the hexagonal unit-cells of energy minimized structures are all given in the supplementary information document. In Table 1, we also compared the lattice energies of predicted monolayers, which reveal decreasing trends in the energetic stability, as the mass of chalcogen atoms in $M_3(C_6X_6)_2$ nanosheets increases. In order to provide a better insight, in Table 1 we also included the lattice parameters of $M_3(C_6N_6H_6)_2$ monolayers take from our earlier study [44]. As it is clear, $M_3(C_6N_6H_6)_2$ and $M_3(C_6O_6)_2$ nanosheets exhibit very close lattice parameters. However, $M_3(C_6O_6)_2$ nanomembranes are energetically distinctly higher stable than $M_3(C_6N_6H_6)_2$ counterparts. Worthy to note that the lattice constants of $Co_3(C_6N_6H_6)_2$, $Ni_3(C_6N_6H_6)_2$ and $Cu_3(C_6O_6)_2$ in the original experimental works were reported to be 13.361 Å [39], 13.113 Å [33] and 13.108 Å [40], respectively, which are very close to the corresponding values we predicted theoretically, 13.40 Å, 13.37 Å and 13.454 Å, respectively.

Table. 1, Lattice parameters of energy minimized $M_3(C_6X_6)_2$ MOF monolayers. Here, $L_{M-X}$, $L_{X-C}$ and $E_{unit-cell}$ stand for metal-chalcogen and chalcogen-C bond lengths and the energy per atom of a unit-cell, respectively. The subscript values belong to the $M_3(C_6N_6H_6)_2$ monolayers taken from our earlier work [44] (X=N).

| Structure | | Lattice constant (Å) | $L_{M-X}$ (Å) | $L_{X-C}$ (Å) | $E_{unit-cell}$ (eV/atom) |
|---|---|---|---|---|---|
| $Co_3C_{12}X_{12}$ | $Co_3C_{12}O_{12}$ | $13.046^{13.40}$ | $1.845^{1.84}$ | $1.304^{1.35}$ | $-7.641^{-7.008}$ |
| | $Co_3C_{12}S_{12}$ | 14.679 | 2.138 | 1.719 | -6.612 |
| | $Co_3C_{12}Se_{12}$ | 15.383 | 2.264 | 1.892 | -6.169 |
| $Cr_3C_{12}X_{12}$ | $Cr_3C_{12}O_{12}$ | $13.458^{13.88}$ | $1.944^{1.98}$ | $1.305^{1.35}$ | $-8.071^{-7.228}$ |
| | $Cr_3C_{12}S_{12}$ | 15.107 | 2.277 | 1.727 | -6.861 |
| | $Cr_3C_{12}Se_{12}$ | 15.780 | 2.388 | 1.906 | -6.413 |
| $Cu_3C_{12}X_{12}$ | $Cu_3C_{12}O_{12}$ | $13.454^{13.78}$ | $1.992^{1.98}$ | $1.274^{1.33}$ | $-7.210^{-6.653}$ |
| | $Cu_3C_{12}S_{12}$ | 14.710 | 2.204 | 1.717 | -6.169 |
| | $Cu_3C_{12}Se_{12}$ | 15.379 | 2.324 | 1.886 | -5.761 |
| $Fe_3C_{12}X_{12}$ | $Fe_3C_{12}O_{12}$ | $13.163^{13.52}$ | $1.852^{1.88}$ | $1.317^{1.35}$ | $-7.728^{-7.095}$ |
| | $Fe_3C_{12}S_{12}$ | 14.806 | 2.177 | 1.724 | -6.748 |
| | $Fe_3C_{12}Se_{12}$ | 15.518 | 2.304 | 1.897 | -6.301 |
| $Mn_3C_{12}X_{12}$ | $Mn_3C_{12}O_{12}$ | $13.251^{13.67}$ | $1.882^{1.92}$ | $1.325^{1.36}$ | $-8.022^{-7.192}$ |
| | $Mn_3C_{12}S_{12}$ | 14.983 | 2.253 | 1.726 | -6.866 |
| | $Mn_3C_{12}Se_{12}$ | 15.677 | 2.383 | 1.895 | -6.419 |
| $Ni_3C_{12}X_{12}$ | $Ni_3C_{12}O_{12}$ | $12.990^{13.37}$ | $1.840^{1.84}$ | $1.304^{1.35}$ | $-7.481^{-6.886}$ |
| | $Ni_3C_{12}S_{12}$ | 14.633 | 2.140 | 1.706 | -6.476 |
| | $Ni_3C_{12}Se_{12}$ | 15.343 | 2.266 | 1.878 | -6.031 |
| $Pd_3C_{12}X_{12}$ | $Pd_3C_{12}O_{12}$ | $13.523^{13.89}$ | $2.004^{2.00}$ | $1.301^{1.34}$ | $-7.341^{-6.806}$ |
| | $Pd_3C_{12}S_{12}$ | 15.074 | 2.284 | 1.706 | -6.406 |
| | $Pd_3C_{12}Se_{12}$ | 15.754 | 2.403 | 1.877 | -5.981 |
| $Rh_3C_{12}X_{12}$ | $Rh_3C_{12}O_{12}$ | $13.599^{13.93}$ | $1.997^{2.00}$ | $1.294^{1.34}$ | $-7.553^{-6.984}$ |
| | $Rh_3C_{12}S_{12}$ | 15.092 | 2.258 | 1.713 | -6.602 |
| | $Rh_3C_{12}Se_{12}$ | 15.768 | 2.370 | 1.895 | -6.171 |

For the application of a material in various devices, presenting acceptable rigidity and strengths are critical for the engineering designs, as they can directly affect the reliability and life service. Such that we first study the mechanical properties of these novel porous 2D lattices by conducting uniaxial tensile simulations. To examine the anisotropy in the mechanical responses of constructed conductive MOFs, in an analogy to graphene the uniaxial tensile simulations were conducted along the armchair and zigzag directions. Worthy to remind that for the uniaxial tensile simulations, the periodic simulation box size along the loading direction was increased gradually. In the same time, the simulation box size along the sheet perpendicular direction of loading was adjusted to reach a negligible stress [44], which is critical to observe the uniaxial stress-conditions. According to our results for the uniaxial stress-strain curves of single-layer $M_3(C_6X_6)_2$(X= O, S, Se), they were found to show similar trends, identical to their $M_3(C_6N_6H_6)_2$ [44] counterparts. In Fig. 2 samples of the DFT predictions for the uniaxial stress-strain responses of single-layer $M_3(C_6X_6)_2$(M= Fe, Rh and X= O, S, Se) elongated along the zigzag and armchair directions are compared. As it is clear, despite of the porous atomic lattices of all considered monolayers, uniaxial stress-strain curves show initial linear relations, corresponding to the linear elasticity. Exhibiting clear linear elasticity in these monolayers reveal that they deform likely to densely packed 2D materials, such as; graphene, $MoS_2$ and borophene [9], in which the stretching from the early stages of loading can be achieved mainly by the bond-elongation rather than bond rotation. It is also clear that for the uniaxial loading along the both considered loading directions, the initial linear parts of the stress-strain relations coincide very closely, revealing convincingly isotropic elasticity in these novel 2D systems. Worthy to remind that within the elastic range, the strain along the traverse direction of loading ($\varepsilon_t$) with respect to the loading strain ($\varepsilon_l$) is constant and can be used to evaluate the Poisson's ratio, using:-$\varepsilon_t/\varepsilon_l$ [44]. In Table 2, the elastic modulus and Poisson's ratio of $M_3(C_6X_6)_2$ monolayers along the armchair and zigzag directions are summarized. Interestingly, by increasing the mass of chalcogen atoms in $M_3(C_6X_6)_2$ nanosheets the elastic modulus decreases. Worthy to note that the elastic modulus of $M_3(C_6O_6)_2$ monolayers are lower than their $M_3(C_6N_6H_6)_2$ [44] counterparts.

According to the results shown in Fig. 2, it is clear that the predicted monolayers exhibit distinctly higher tensile strengths and strain at tensile strength point (stretchability) along the armchair than the zigzag direction. Such trends were also found to be consistent for the most of considered $M_3(C_6X_6)_2$ nanosheets, however for few cases the tensile strengths and stretchability were found to be considerably close along the both loading directions. In Table 2, the tensile strength and stretchability of studied monolayers are also compared. Notably, $M_3(C_6X_6)_2$ nanosheets with porous atomic lattices can yield comparable or even higher their elastic modulus and tensile strengths than some densely packed 2D materials, like; germanene and stanene [55]. It is also worth noting that according to the Griffth theory [56], the tensile strength of a densely packed material is around one tenth of the elastic modulus ($UTS \approx E/10$). We remind that this ratio, for pristine graphene, $MoS_2$, $MoTe_2$, bucked borophene and silicene were reported to be, 8 [3], ~11 [57], ~7-10 [58], ~12-17 [59] and ~9 [55], respectively. As compared in the Table 1, in this regard the $M_3(C_6X_6)_2$ nanomembranes also behave like densely packed 2D materials. Notably, in the cases of $Ni_3$- and $Pd_3$-$C_{12}O_{12}$ the ratio between the elastic modulus and tensile strength are over 16 and 14, respectively.

With respect to the deformation/failure mechanism, we observed identical behaviour for the all studied monolayers, exactly similar to what we found in our work for the $M_3(C_6N_6H_6)_2$ nanosheets [44]. Just as examples, in Fig. 3 we illustrated the top views of $Fe_3(C_6X_6)_2$(X= O, S, Se) monolayers at strain levels shortly after the tensile strength elongated along the armchair and zigzag directions. As a general rule for the all $M_3(C_6X_6)_2$(X= NH, O, S, Se) nanosheets stretched along the different loading directions, the failure always initiates by the breakage of metal-X bonds. This finding confirms that the bonding characteristics of metal-X bonds can dominate the mechanical properties of $M_3(C_6X_6)_2$ nanosheets.

We also examined the thermal stability of predicted 2D MOFs by conducting the AIMD simulations for 15 ps. According to our results (shown in Fig. S1), all the constructed nanomembranes could stay intact at the high temperature of 1500 K, which confirm their remarkable thermal stability. The maximum variations in the bond lengths were found to occur in the cases of metal-X bonds, which can be easily explained because of the fact that they are originally more elongated than the other bonds in these systems. As discussed earlier, metal-X bonds are also the softest bonds in these nanosheets as the failures always initiate by the breakage of these bonds. Worth noting that because of the tremendous computational costs of AIMD simulations, in this work we did not find and compare the temperatures that these novel nanosheets disintegrate. In addition, analysis of root-mean-square deviation of atomic positions during the AIMD simulations can provide useful vision concerning the dynamical response of predicted nanosheets. Therefore the investigation of these effects can be interesting topics for the future works.

Table 2, Summarized mechanical properties of single-layer $M_3(C_6X_6)_2$ elongated along the armchair and zigzag directions. *E*, *P*, *UTS* and *SUTS* stand for elastic modulus, Poisson's ratio, ultimate tensile strength and strain at ultimate tensile strength point, respectively.

| Structure | Direction | *E* (N/m) | *P* | *UTS* (N/m) | *SUTS* | *E/UTS* |
|---|---|---|---|---|---|---|
| $Co_3C_{12}O_{12}$ | Armchair | 41 | 0.42 | 4.1 | 0.12 | 10 |
| | Zigzag | 41 | 0.42 | 3.6 | 0.11 | 11.34 |
| $Co_3C_{12}S_{12}$ | Armchair | 31 | 0.4 | 4.2 | 0.18 | 7.4 |
| | Zigzag | 33 | 0.4 | 3.8 | 0.18 | 8.7 |
| $Co_3C_{12}Se_{12}$ | Armchair | 28 | 0.37 | 3.6 | 0.19 | 7.8 |
| | Zigzag | 28 | 0.37 | 3.2 | 0.19 | 8.8 |
| $Cr_3C_{12}O_{12}$ | Armchair | 32 | 0.41 | 3.7 | 0.16 | 8.6 |
| | Zigzag | 32 | 0.41 | 2.7 | 0.14 | 11.9 |
| $Cr_3C_{12}S_{12}$ | Armchair | 26 | 0.33 | 3.1 | 0.18 | 8.4 |
| | Zigzag | 26 | 0.31 | 2.7 | 0.18 | 9.6 |
| $Cr_3C_{12}Se_{12}$ | Armchair | 24 | 0.32 | 2.7 | 0.19 | 8.9 |
| | Zigzag | 23 | 0.31 | 2.3 | 0.18 | 10 |
| $Cu_3C_{12}S_{12}$ | Armchair | 26 | 0.38 | 2.8 | 0.17 | 9.3 |
| | Zigzag | 26 | 0.40 | 2.2 | 0.14 | 11.8 |
| $Cu_3C_{12}Se_{12}$ | Armchair | 23 | 0.37 | 2.7 | 0.19 | 8.5 |
| | Zigzag | 22 | 0.34 | 2.3 | 0.16 | 9.6 |
| $Fe_3C_{12}O_{12}$ | Armchair | 30 | 0.41 | 4.0 | 0.18 | 7.5 |
| | Zigzag | 30 | 0.39 | 3.6 | 0.17 | 8.3 |
| $Fe_3C_{12}S_{12}$ | Armchair | 30 | 0.4 | 4.0 | 0.19 | 7.5 |
| | Zigzag | 30 | 0.4 | 3.5 | 0.18 | 8.6 |
| $Fe_3C_{12}Se_{12}$ | Armchair | 27 | 0.35 | 3.4 | 0.19 | 7.9 |
| | Zigzag | 27 | 0.35 | 3.0 | 0.18 | 9 |
| $Mn_3C_{12}O_{12}$ | Armchair | 39 | 0.4 | 3.0 | 0.085 | 13 |
| | Zigzag | 39 | 0.42 | 3.1 | 0.09 | 12.6 |
| $Mn_3C_{12}S_{12}$ | Armchair | 27 | 0.29 | 3.4 | 0.19 | 7.9 |
| | Zigzag | 27 | 0.32 | 3.0 | 0.19 | 9 |
| $Mn_3C_{12}Se_{12}$ | Armchair | 24 | 0.31 | 2.9 | 0.2 | 8.3 |
| | Zigzag | 24 | 0.27 | 2.5 | 0.19 | 9.6 |
| $Ni_3C_{12}O_{12}$ | Armchair | 42 | 0.38 | 2.5 | 0.07 | 16.8 |
| | Zigzag | 40 | 0.38 | 2.5 | 0.075 | 16 |
| $Ni_3C_{12}S_{12}$ | Armchair | 31 | 0.4 | 4.2 | 0.19 | 7.4 |
| | Zigzag | 30 | 0.38 | 3.6 | 0.19 | 8.3 |
| $Ni_3C_{12}Se_{12}$ | Armchair | 26 | 0.39 | 3.6 | 0.19 | 7.2 |
| | Zigzag | 25 | 0.36 | 3.3 | 0.2 | 7.6 |
| $Pd_3C_{12}O_{12}$ | Armchair | 34 | 0.4 | 2.4 | 0.08 | 14.2 |
| | Zigzag | 34 | 0.4 | 2.4 | 0.09 | 14.2 |
| $Pd_3C_{12}S_{12}$ | Armchair | 29 | 0.32 | 3.6 | 0.18 | 8.1 |
| | Zigzag | 28 | 0.32 | 3.0 | 0.19 | 9.3 |
| $Pd_3C_{12}Se_{12}$ | Armchair | 25 | 0.31 | 3.1 | 0.2 | 8.1 |
| | Zigzag | 25 | 0.28 | 2.6 | 0.21 | 9.6 |
| $Rh_3C_{12}O_{12}$ | Armchair | 36 | 0.43 | 4.2 | 0.17 | 8.6 |
| | Zigzag | 36 | 0.41 | 3.0 | 0.15 | 12 |
| $Rh_3C_{12}S_{12}$ | Armchair | 31 | 0.37 | 4.3 | 0.19 | 7.2 |
| | Zigzag | 31 | 0.37 | 3.6 | 0.17 | 8.6 |
| $Rh_3C_{12}Se_{12}$ | Armchair | 27 | 0.37 | 3.7 | 0.2 | 7.3 |
| | Zigzag | 27 | 0.34 | 3.2 | 0.18 | 8.4 |

We next investigate the electronic and magnetic properties of $M_3(C_6X_6)_2$ monolayers. To this aim, electronic density of states (DOS) were calculated for the both spin up and down channels and the acquired results are illustrated Figs. 4, 5 and 6, respectively, for $M_3(C_6O_6)_2$, $M_3(C_6S_6)_2$ and $M_3(C_6Se_6)_2$ monolayers. In these results the energy values are plotted with respect to the Fermi energy ($E$-$E_f$), in which the valance band maximum occurs at the Fermi level. For all $M_3(C_6O_6)_2$ nanosheets the spin splitting of DOS around Fermi energy is occurring, which is basically due to the strong hybridization of transition metal 3d orbitals with s and p orbitals of O atoms. There is also weak hybridization of transition metal 3d orbitals and oxygen orbitals with 2s and 2p orbitals of carbon atoms. For the all predicted nanomembranes, our calculations reveal that the ferromagnetic (FM) phase is more stable than nonmagnetic (NM) phase. Our results indicate that the $Cr_3$-, $Co_3$-, $Mn_3$- and $Pd_3$-$(C_6O_6)_2$ monolayers are half-metallic ferromagnetic systems with 100% spin polarization, because of the fact that in these systems the valence band for one spin orientation is partially filled, while there is a gap in the density of states for the other spin orientation (Fig. 5). The spin polarization, $P$, is defined as follows:

$$P = \frac{N_{Ef}^{\uparrow} - N_{Ef}^{\downarrow}}{N_{Ef}^{\uparrow} + N_{Ef}^{\downarrow}} \quad (1)$$

where $N_{Ef}^{\uparrow}$ and $N_{Ef}^{\downarrow}$ represent the density of states of majority spin (spin up) and minority spin (spin down) at the Fermi level, respectively. The spin-flip gap which is important for the electron injection in half metals and can be defined as the separation of the Fermi level from the minimum of the conduction band in minority spin, was found to be 0.00, 0.00, 0.06 and 0.00 eV for $Cr_3$-, $Co_3$-, $Mn_3$- and $Pd_3$-$(C_6O_6)_2$ monolayers, respectively. The $Cu_3$- and $Fe_3$-$(C_6O_6)_2$ monolayers were found to be magnetic semiconductors, since there is a gap in the density of states for the both spin channels. We also calculated entire magnetic moment and partial magnetic moments of each magnetic atom for these monolayers. The magnetization, $M$ is defined as follows:

$$M = (N^{\uparrow} - N^{\downarrow}) \times \mu_B \quad (2)$$

where $N^{\uparrow}$ and $N^{\downarrow}$ stand for the total number of electrons of majority spin and total number of electrons of minority spin, respectively. The total magnetic moment per magnetic atoms was measured to be 2.75, 0.54, 2.95, 0.36, 0.90 and 2.14 $\mu_B$ for the $Cr_3$-, $Co_3$-, $Mn_3$-, $Pd_3$-, $Cu_3$- and $Fe_3$-$(C_6O_6)_2$ monolayers, respectively. The $Ni_3$- and $Rh_3$-$(C_6O_6)_2$ nanosheets show metallic features for the both spin up and down channels with 28% and 19.28% spin polarization, respectively. The total magnetic moment for these monolayers is 0.19 and 0.17 $\mu_B$, respectively. In Table 3 the electronic and magnetic properties of predicted monolayers are summarized.

Results shown in Fig. 5 suggest that $Co_3$- $Fe_3$- and $Cr_3$-$(C_6S_6)_2$ MOFs systems are magnetic semiconductors without spin polarization, whereas $Mn_3(C_6S_6)_2$ monolayers are half-metallic systems with 100% spin polarization. The spin-flip gap was measured to be 0.43 eV for $Mn_3$-$(C_6S_6)_2$ nanosheets. Also $Cu_3$- and $Rh_3$-$(C_6S_6)_2$ monolayers show metallic features for both spin channels with 95% and 12.5% spin polarization, respectively. The total magnetic moment per magnetic atoms for $Co_3$-, $Cr_3$-, $Mn_3$-, $Cu_3$- and $Fe_3$-$(C_6S_6)_2$ monolayers was found to be 0.97, 2.17, 2.92, 0.25 and 1.96 $\mu_B$, respectively. Among $M_3(C_6S_6)_2$ nanosheets the $Pd_3$- and $Ni_3$-$(C_6S_6)_2$ show nonmagnetic metallic character.

Table 3, Calculated values of the spin polarization, total magnetic moment per magnetic atom (TMM) and partial magnetic moments of the magnetic atom (PMM) (in Bohr magnetons $\mu_B$) of $M_3(C_6X_6)_2$ monolayers.

| Structure | Polarization (%) | TMM | PMM |
|---|---|---|---|
| $Cr_3(C_6O_6)_2$ | 100 | 2.75 | 3.20 |
| $Co_3(C_6O_6)_2$ | 100 | 0.54 | 0.90 |
| $Cu_3(C_6O_6)_2$ | 0.00 | 0.90 | 0.45 |
| $Fe_3(C_6O_6)_2$ | 0.00 | 2.14 | 2.18 |
| $Mn_3(C_6O_6)_2$ | 100 | 2.95 | 3.20 |
| $Ni_3(C_6O_6)_2$ | 28 | 0.19 | 0.03 |
| $Pd_3(C_6O_6)_2$ | 100 | 0.36 | 0.01 |
| $Rh_3(C_6O_6)_2$ | 19.28 | 0.17 | 0.11 |
| $Cr_3(C_6S_6)_2$ | 0.00 | 2.17 | 2.85 |
| $Co_3(C_6S_6)_2$ | 0.00 | 0.97 | 1.00 |
| $Cu_3(C_6S_6)_2$ | 95 | 0.25 | 0.10 |
| $Fe_3(C_6S_6)_2$ | 0.00 | 1.96 | 2.17 |
| $Mn_3(C_6S_6)_2$ | 100 | 2.92 | 3.20 |
| $Ni_3(C_6S_6)_2$ | 0.00 | 0.00 | 0.00 |
| $Pd_3(C_6S_6)_2$ | 0.00 | 0.00 | 0.00 |
| $Rh_3(C_6S_6)_2$ | 12.50 | 0.42 | 0.30 |
| $Cr_3(C_6Se_6)_2$ | 100 | 2.17 | 2.85 |
| $Co_3(C_6Se_6)_2$ | 0.00 | 1.00 | 1.06 |
| $Cu_3(C_6Se_6)_2$ | 0.00 | 0.20 | 0.02 |
| $Fe_3(C_6Se_6)_2$ | 0.00 | 2.00 | 2.27 |
| $Mn_3(C_6Se_6)_2$ | 100 | 2.96 | 3.41 |
| $Ni_3(C_6Se_6)_2$ | 0.00 | 0.00 | 0.00 |
| $Pd_3(C_6Se_6)_2$ | 0.00 | 0.00 | 0.00 |
| $Rh_3(C_6Se_6)_2$ | 0.00 | 0.77 | 0.51 |

Fig. 6 shows the density of states for $M_3(C_6Se_6)_2$ monolayers. Our results show that among all $M_3(C_6Se_6)_2$ monolayers the $Co_3$-, $Fe_3$- and $Rh_3(C_6Se_6)_2$ structures are magnetic semiconductors with the total magnetic moment of 1.00, 2.00, and 0.77 $\mu_{B,}$ respectively. $Cr_3$- and $Mn_3$-$(C_6Se_6)_2$ monolayers are half-metallic systems with 100% spin polarization with the spin-flip gap of 0.89 and 0.48 eV and magnetic moment of 2.17 and 2.96 $\mu_B$, respectively. Moreover only $Cu_3(C_6Se_6)_2$ monolayer is magnetic metals with 0.00% spin polarization and 0.20 $\mu_B$ magnetization since the DOS in the both channels crossing the Fermi level, hence it does not have the perfect spin polarization. The two other systems, $Ni_3$- and $Pd_3$-$(C_6Se_6)_2$ monolayers, are nonmagnetic metals without spin polarization and magnetic moment. To further discern the electronic feature of these monolayers, the partial DOS for $Cr_3(C_6X_6)_2$ and $Fe_3(C_6X_6)_2$ structures are shown in Fig. 7. As It can be seen for $Fe_3$- and $Cr_3(C_6X_6)_2$ nanosheets, the spin splitting of DOS around the Fermi energy is occurring, which is basically due to the strong hybridization of transition metal 3d orbitals with s and p orbitals of X (O, S and Se) atoms. Weak hybridization of transition metal 3d orbitals and oxygen orbitals with 2s and 2p orbitals of carbon atoms can be also observed. Similar trends were also found to exist for the other studied nanosheets. In comparison with $M_3(C_6N_6H_6)_2$ nanosheets explored in our previous work [44], the $M_3(C_6X_6)_2$(X= O, S, Se) counterparts are more diverse in terms of electronic and magnetic properties. It was found that $M_3(C_6N_6H_6)_2$ monolayers show perfect half metallic and nonmagnetic metallic behavior while $M_3(C_6X_6)_2$(X= O, S, Se) MOFs act as magnetic semiconducting, perfect half metallic, magnetic and nonmagnetic metallic systems. The high magnetic moment and spin polarization of $M_3(C_6X_6)_2$(X= O, S, Se) nanosheets suggest that they can be used in electronic devices, magnetic recording and spintronic applications, e.g. spin-transistors.

## 4. Concluding remarks

Conductive 2D metal–organic frameworks (MOFs) are among the most attractive classes of materials, with excellent application prospects for the energy storage, electrocatalysis and electronic systems. Most recently, Cu-hexahydroxybenzene, a novel 2D and conductive MOF with a chemical formula of $Cu_3(C_6O_6)_2$ was synthesized, with excellent application prospects for nanoelectronics, sensing and energy-related systems. This experimental advance highlights that whether other $M_3(C_6X_6)_2$ conductive MOFs nanosheets with different combinations of metal (M) and chalcogen (X) atoms can be stable and synthesizable? and moreover, how attractive would be their intrinsic properties?. In this short communication, we accordingly investigated thermal stability, mechanical properties and electronic characteristics of $M_3(C_6X_6)_2$(M= Co, Cr, Cu, Fe, Mn, Ni, Pd, Rh and X= O, S, Se) monolayers using the first-principles density functional theory calculations. It was confirmed that these novel porous nanosheets can exhibit linear elasticity with remarkable tensile strengths, comparable to some other densely packed 2D materials. It was shown that by increasing the mass of chalcogen atoms in $M_3(C_6X_6)_2$ nanosheets, the elastic modulus decreases. Analysis of deformation process confirmed that the characteristics of metal-chalcogen bonds dominate the mechanical/failure responses of these novel 2D systems. Ab-initio molecular dynamics simulations confirm the thermal stability of all predicted monolayers, with the intact atomic lattices at the high temperature of 1500 K. Electronic and magnetic properties of $M_3(C_6X_6)_2$ conductive MOFs monolayers were explored using DFT within the PBE approximation. Our results illustrate that for different combination of metal and chalcogen atoms these monolayers can yield; magnetic semiconductor, perfect half metal, magnetic and nonmagnetic metal systems. The high magnetic moment and spin polarization of some of these nanosheets imply that they might be suitable for the use in the applications related to; permanent magnetism, magnetic recording, electronic and spintronic nanodevices. Nevertheless, as it has been experimentally proven, conductive MOFs with porous atomic lattices may serve as highly desirable candidates for the design of advanced energy storage/conversion systems.

**Acknowledgment**

B. M. and T. R. greatly acknowledge the financial support by European Research Council for COMBAT project (Grant number 615132). T. H. is indebted to the resources at NCI National Facility systems at the Australian National University.

**Data availability**

The energy minimized atomic lattices in the VASP POSCAR format and PAW potentials in the VASP POTCAR format are available to download from:

**References**

[1] K.S. Novoselov, A.K. Geim, S. V Morozov, D. Jiang, Y. Zhang, S. V Dubonos, I. V Grigorieva, A.A. Firsov, Electric field effect in atomically thin carbon films., Science. 306 (2004) 666–9. doi:10.1126/science.1102896.

[2] A.K. Geim, K.S. Novoselov, The rise of graphene, Nat. Mater. 6 (2007) 183–191. doi:10.1038/nmat1849.

[3] C. Lee, X. Wei, J.W. Kysar, J. Hone, Measurement of the Elastic Properties and Intrinsic Strength of Monolayer Graphene, Science (80-. ). 321 (2008) 385–388. doi:10.1126/science.1157996.

[4] A.A. Balandin, S. Ghosh, W. Bao, I. Calizo, D. Teweldebrhan, F. Miao, C.N. Lau, Superior thermal conductivity of single-layer graphene, Nano Lett. 8 (2008) 902–907. doi:10.1021/nl0731872.

[5] A.A. Balandin, Thermal properties of graphene and nanostructured carbon materials, Nat. Mater. 10 (2011) 569–581. doi:10.1038/nmat3064.

[6] C. Berger, Z. Song, T. Li, X. Li, A.Y. Ogbazghi, R. Feng, Z. Dai, A.N. Marchenkov, E.H. Conrad, P.N. First, W. a de Heer, Ultrathin Epitaxial Graphite:  2D Electron Gas Properties and a Route toward Graphene-based Nanoelectronics, J. Phys. Chem. B. 108 (2004) 19912–19916. doi:doi:10.1021/jp040650f.

[7] M. Liu, X. Yin, E. Ulin-Avila, B. Geng, T. Zentgraf, L. Ju, F. Wang, X. Zhang, A graphene-based broadband optical modulator, Nature. 474 (2011) 64–67. doi:10.1038/nature10067.

[8] F. Withers, M. Dubois, A.K. Savchenko, Electron properties of fluorinated single-layer graphene transistors, Phys. Rev. B - Condens. Matter Mater. Phys. 82 (2010). doi:10.1103/PhysRevB.82.073403.

[9] B. Liu, K. Zhou, Recent progress on graphene-analogous 2D nanomaterials:

Properties, modeling and applications, Prog. Mater. Sci. 100 (2019) 99–169. doi:10.1016/J.PMATSCI.2018.09.004.
[10] L. Wang, M. Pumera, Electrochemical catalysis at low dimensional carbons: Graphene, carbon nanotubes and beyond – A review, Appl. Mater. Today. 5 (2016) 134–141. doi:10.1016/j.apmt.2016.09.011.
[11] B. Radisavljevic, A. Radenovic, J. Brivio, V. Giacometti, A. Kis, Single-layer MoS2 transistors, Nat. Nanotechnol. 6 (2011) 147–50. doi:10.1038/nnano.2010.279.
[12] Q.H. Wang, K. Kalantar-Zadeh, A. Kis, J.N. Coleman, M.S. Strano, Electronics and optoelectronics of two-dimensional transition metal dichalcogenides, Nat. Nanotechnol. 7 (2012) 699–712. doi:10.1038/nnano.2012.193.
[13] A. Eftekhari, Molybdenum diselenide (MoSe2) for energy storage, catalysis, and optoelectronics, Appl. Mater. Today. 8 (2017) 1–17. doi:http://dx.doi.org/10.1016/j.apmt.2017.01.006.
[14] Y. Wang, Z. Sofer, J. Luxa, M. Pumera, Lithium Exfoliated Vanadium Dichalcogenides (VS2, VSe2, VTe2) Exhibit Dramatically Different Properties from Their Bulk Counterparts, Adv. Mater. Interfaces. 3 (2016). doi:10.1002/admi.201600433.
[15] S. Presolski, M. Pumera, Covalent functionalization of MoS2, Mater. Today. 19 (2016) 140–145. doi:10.1016/j.mattod.2015.08.019.
[16] S. Das, M. Demarteau, A. Roelofs, Ambipolar phosphorene field effect transistor, ACS Nano. 8 (2014) 11730–11738. doi:10.1021/nn505868h.
[17] L. Li, Y. Yu, G.J. Ye, Q. Ge, X. Ou, H. Wu, D. Feng, X.H. Chen, Y. Zhang, Black phosphorus field-effect transistors, Nat. Nanotechnol. 9 (2014) 372–377. doi:10.1038/nnano.2014.35.
[18] A. Thomas, A. Fischer, F. Goettmann, M. Antonietti, J.-O. Müller, R. Schlögl, J.M. Carlsson, Graphitic carbon nitride materials: variation of structure and morphology and their use as metal-free catalysts, J. Mater. Chem. 18 (2008) 4893. doi:10.1039/b800274f.
[19] G. Algara-Siller, N. Severin, S.Y. Chong, T. Björkman, R.G. Palgrave, A. Laybourn, M. Antonietti, Y.Z. Khimyak, A. V. Krasheninnikov, J.P. Rabe, U. Kaiser, A.I. Cooper, A. Thomas, M.J. Bojdys, Triazine-based graphitic carbon nitride: A two-dimensional semiconductor, Angew. Chemie - Int. Ed. 53 (2014) 7450–7455. doi:10.1002/anie.201402191.
[20] J. Mahmood, E.K. Lee, M. Jung, D. Shin, I.-Y. Jeon, S.-M. Jung, H.-J. Choi, J.-M. Seo, S.-Y. Bae, S.-D. Sohn, N. Park, J.H. Oh, H.-J. Shin, J.-B. Baek, Nitrogenated holey two-dimensional structures, Nat. Commun. 6 (2015) 6486. doi:10.1038/ncomms7486.
[21] L.-B. Shi, Y.-Y. Zhang, X.-M. Xiu, H.-K. Dong, Structural, electronic and adsorptive characteristics of phosphorated holey graphene (PHG): First principles calculations, Diam. Relat. Mater. 82 (2018) 102–108. doi:https://doi.org/10.1016/j.diamond.2018.01.004.
[22] L.-B. Shi, Y.-Y. Zhang, X.-M. Xiu, H.-K. Dong, Structural characteristics and strain behaviors of two-dimensional C3N : First principles calculations, Carbon N. Y. 134 (2018) 103–111. doi:https://doi.org/10.1016/j.carbon.2018.03.076.
[23] L. Bin Shi, S. Cao, J. Zhang, X.M. Xiu, H.K. Dong, Mechanical behaviors and electronic characteristics on two-dimensional C2N3and C2N3H: First principles calculations, Phys. E Low-Dimensional Syst. Nanostructures. 103 (2018) 252–263. doi:10.1016/j.physe.2018.06.014.
[24] H.L. Lee, Z. Sofer, V. Mazánek, J. Luxa, C.K. Chua, M. Pumera, Graphitic carbon nitride:

Effects of various precursors on the structural, morphological and electrochemical sensing properties, Appl. Mater. Today. 8 (2017) 150–162. doi:10.1016/j.apmt.2016.09.019.
[25] R.H. Baughman, H. Eckhardt, M. Kertesz, Structure-property predictions for new planar forms of carbon: Layered phases containing $sp^2$ and sp atoms, J. Chem. Phys. 87 (1987) 6687. doi:10.1063/1.453405.
[26] G. Li, Y. Li, H. Liu, Y. Guo, Y. Li, D. Zhu, Architecture of graphdiyne nanoscale films, Chem. Commun. 46 (2010) 3256–3258. doi:10.1039/B922733D.
[27] X. Kan, Y. Ban, C. Wu, Q. Pan, H. Liu, J. Song, Z. Zuo, Z. Li, Y. Zhao, Interfacial Synthesis of Conjugated Two-Dimensional N-Graphdiyne, ACS Appl. Mater. Interfaces. 10 (2018) 53–58. doi:10.1021/acsami.7b17326.
[28] N. Wang, X. Li, Z. Tu, F. Zhao, J. He, Z. Guan, C. Huang, Y. Yi, Y. Li, Synthesis, Electronic Structure of Boron-Graphdiyne with an sp-Hybridized Carbon Skeleton and Its Application in Sodium Storage, Angew. Chemie. (2018). doi:10.1002/ange.201801897.
[29] H. Shang, Z. Zuo, L. Li, F. Wang, H. Liu, Y. Li, Y. Li, Ultrathin Graphdiyne Nanosheets Grown In Situ on Copper Nanowires and Their Performance as Lithium-Ion Battery Anodes, Angew. Chemie Int. Ed. 57 (2017) 774–778. doi:10.1002/anie.201711366.
[30] H. Shang, Z. Zuo, L. Yu, F. Wang, F. He, Y. Li, Low-Temperature Growth of All-Carbon Graphdiyne on a Silicon Anode for High-Performance Lithium-Ion Batteries, Adv. Mater. 30 (2018) 1801459. doi:10.1002/adma.201801459.
[31] R. Dong, M. Pfeffermann, H. Liang, Z. Zheng, X. Zhu, J. Zhang, X. Feng, Large-Area, Free-Standing, Two-Dimensional Supramolecular Polymer Single-Layer Sheets for Highly Efficient Electrocatalytic Hydrogen Evolution, Angew. Chemie - Int. Ed. (2015). doi:10.1002/anie.201506048.
[32] T. Kambe, R. Sakamoto, K. Hoshiko, K. Takada, M. Miyachi, J.H. Ryu, S. Sasaki, J. Kim, K. Nakazato, M. Takata, H. Nishihara, π-Conjugated nickel bis(dithiolene) complex nanosheet, J. Am. Chem. Soc. (2013). doi:10.1021/ja312380b.
[33] D. Feng, T. Lei, M.R. Lukatskaya, J. Park, Z. Huang, M. Lee, L. Shaw, S. Chen, A.A. Yakovenko, A. Kulkarni, J. Xiao, K. Fredrickson, J.B. Tok, X. Zou, Y. Cui, Z. Bao, Robust and conductive two-dimensional metal-organic frameworks with exceptionally high volumetric and areal capacitance, Nat. Energy. 3 (2018) 30–36. doi:10.1038/s41560-017-0044-5.
[34] H. Nagatomi, N. Yanai, T. Yamada, K. Shiraishi, N. Kimizuka, Synthesis and Electric Properties of a Two-Dimensional Metal-Organic Framework Based on Phthalocyanine, Chem. - A Eur. J. (2018). doi:10.1002/chem.201705530.
[35] M.G. Campbell, D. Sheberla, S.F. Liu, T.M. Swager, M. Dincə, Cu3(hexaiminotriphenylene)2: An electrically conductive 2D metal-organic framework for chemiresistive sensing, Angew. Chemie - Int. Ed. 54 (2015) 4349–4352. doi:10.1002/anie.201411854.
[36] M. Ko, A. Aykanat, M.K. Smith, K.A. Mirica, Drawing sensors with ball-milled blends of metal-organic frameworks and graphite, Sensors (Switzerland). (2017). doi:10.3390/s17102192.
[37] M.K. Smith, K.A. Mirica, Self-Organized Frameworks on Textiles (SOFT): Conductive Fabrics for Simultaneous Sensing, Capture, and Filtration of Gases, J. Am. Chem. Soc. 139 (2017) 16759–16767. doi:10.1021/jacs.7b08840.
[38] J. Park, M. Xu, F. Li, H.C. Zhou, 3D Long-Range Triplet Migration in a Water-Stable Metal-Organic Framework for Upconversion-Based Ultralow-Power in Vivo Imaging, J.

Am. Chem. Soc. 140 (2018) 5493–5499. doi:10.1021/jacs.8b01613.
[39] J. Park, M. Lee, D. Feng, Z. Huang, A.C. Hinckley, A. Yakovenko, X. Zou, Y. Cui, Z. Bao, Stabilization of Hexaaminobenzene in a 2D Conductive Metal-Organic Framework for High Power Sodium Storage, J. Am. Chem. Soc. 140 (2018) 10315–10323. doi:10.1021/jacs.8b06020.
[40] J. Park, A.C. Hinckley, Z. Huang, D. Feng, A. Yakovenko, M. Lee, S. Chen, X. Zou, Z. Bao, Synthetic Routes for a 2D Semiconductive Copper Hexahydroxybenzene Metal–Organic Framework, J. Am. Chem. Soc. 140 (2018) 14533–14537. doi:10.1021/jacs.8b06666.
[41] A.R. Oganov, C.W. Glass, Crystal structure prediction using ab initio evolutionary techniques: principles and applications., J. Chem. Phys. 124 (2006) 244704. doi:10.1063/1.2210932.
[42] C.W. Glass, A.R. Oganov, N. Hansen, USPEX-Evolutionary crystal structure prediction, Comput. Phys. Commun. 175 (2006) 713–720. doi:10.1016/j.cpc.2006.07.020.
[43] L.-B. Shi, S. Cao, M. Yang, Strain behavior and Carrier mobility for novel two-dimensional semiconductor of GeP: First principles calculations, Phys. E Low-Dimensional Syst. Nanostructures. 107 (2019) 124–130. doi:10.1016/J.PHYSE.2018.11.024.
[44] B. Mortazavi, M. Shahrokhi, M. Makaremi, G. Cuniberti, T. Rabczuk, First-principles investigation of Ag-, Co-, Cr-, Cu-, Fe-, Mn-, Ni-, Pd- and Rh-hexaaminobenzene 2D metal-organic frameworks, Mater. Today Energy. 10 (2018) 336–342. doi:10.1016/J.MTENER.2018.10.007.
[45] Y. Liu, X. Peng, Recent advances of supercapacitors based on two-dimensional materials, Appl. Mater. Today. 7 (2017) 1–12. doi:http://dx.doi.org/10.1016/j.apmt.2017.01.004.
[46] G. Kresse, J. Furthmüller, Efficiency of ab-initio total energy calculations for metals and semiconductors using a plane-wave basis set, Comput. Mater. Sci. 6 (1996) 15–50. doi:10.1016/0927-0256(96)00008-0.
[47] G. Kresse, J. Furthmüller, Efficient iterative schemes for ab initio total-energy calculations using a plane-wave basis set, Phys. Rev. B. 54 (1996) 11169–11186. doi:10.1103/PhysRevB.54.11169.
[48] G. Kresse, From ultrasoft pseudopotentials to the projector augmented-wave method, Phys. Rev. B. 59 (1999) 1758–1775. doi:10.1103/PhysRevB.59.1758.
[49] J. Perdew, K. Burke, M. Ernzerhof, Generalized Gradient Approximation Made Simple., Phys. Rev. Lett. 77 (1996) 3865–3868. doi:10.1103/PhysRevLett.77.3865.
[50] P.E. Blöchl, Projector augmented-wave method, Phys. Rev. B. 50 (1994) 17953–17979. doi:10.1103/PhysRevB.50.17953.
[51] K. Momma, F. Izumi, VESTA 3 for three-dimensional visualization of crystal, volumetric and morphology data, J. Appl. Crystallogr. 44 (2011) 1272–1276. doi:10.1107/S0021889811038970.
[52] H. Monkhorst, J. Pack, Special points for Brillouin zone integrations, Phys. Rev. B. 13 (1976) 5188–5192. doi:10.1103/PhysRevB.13.5188.
[53] F. Herman, C.D. Kuglin, K.F. Cuff, R.L. Kortum, Relativistic corrections to the band structure of tetrahedrally bonded semiconductors, Phys. Rev. Lett. (1963). doi:10.1103/PhysRevLett.11.541.
[54] B. Silvi, A. Savin, Classification of Chemical-Bonds Based on Topological Analysis of Electron Localization Functions, Nature. 371 (1994) 683–686. doi:10.1038/371683a0.

[55] B. Mortazavi, O. Rahaman, M. Makaremi, A. Dianat, G. Cuniberti, T. Rabczuk, First-principles investigation of mechanical properties of silicene, germanene and stanene, Phys. E Low-Dimensional Syst. Nanostructures. 87 (2017) 228–232. doi:10.1016/j.physe.2016.10.047.
[56] A.A. Griffith, The Phenomena of Rupture and Flow in Solids, Philos. Trans. R. Soc. A Math. Phys. Eng. Sci. (1921). doi:10.1098/rsta.1921.0006.
[57] S. Bertolazzi, J. Brivio, A. Kis, Stretching and breaking of ultrathin MoS 2, ACS Nano. 5 (2011) 9703–9709. doi:10.1021/nn203879f.
[58] B. Mortazavi, G.R. Berdiyorov, M. Makaremi, T. Rabczuk, Mechanical responses of two-dimensional MoTe2; pristine 2H, 1T and 1T′ and 1T′/2H heterostructure, Extrem. Mech. Lett. 20 (2018) 65–72. doi:10.1016/J.EML.2018.01.005.
[59] B. Mortazavi, O. Rahaman, A. Dianat, T. Rabczuk, Mechanical responses of borophene sheets: A first-principles study, Phys. Chem. Chem. Phys. 18 (2016) 27405–27413. doi:10.1039/C6CP03828J.

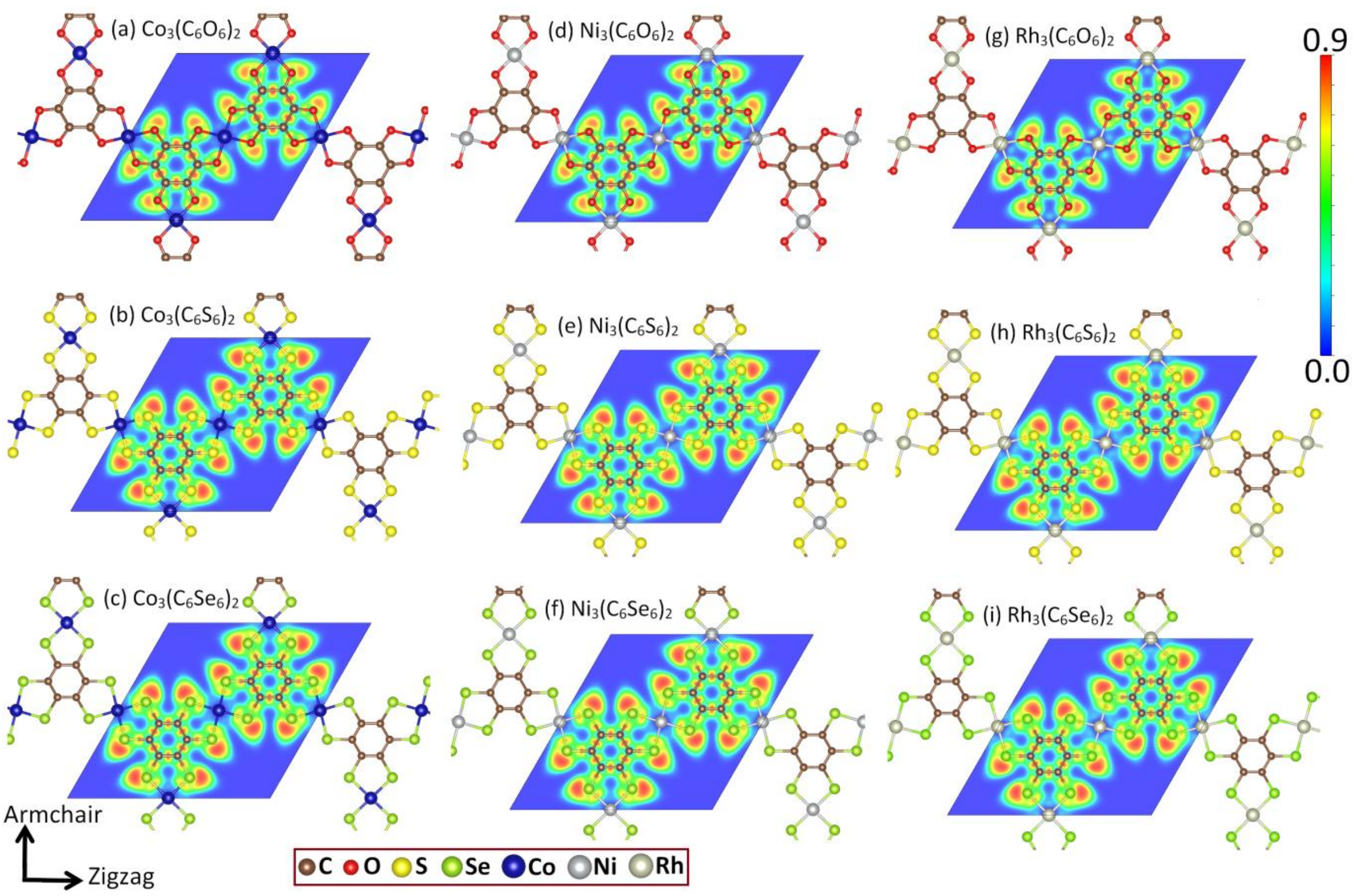

Fig. 1, Samples of atomic structures of energy minimized $M_3(C_6X_6)_2$(M=Co, Ni, Rh and X=O, S, Se) monolayers. Contours illustrate the electron localization function [54] within the unit-cell.

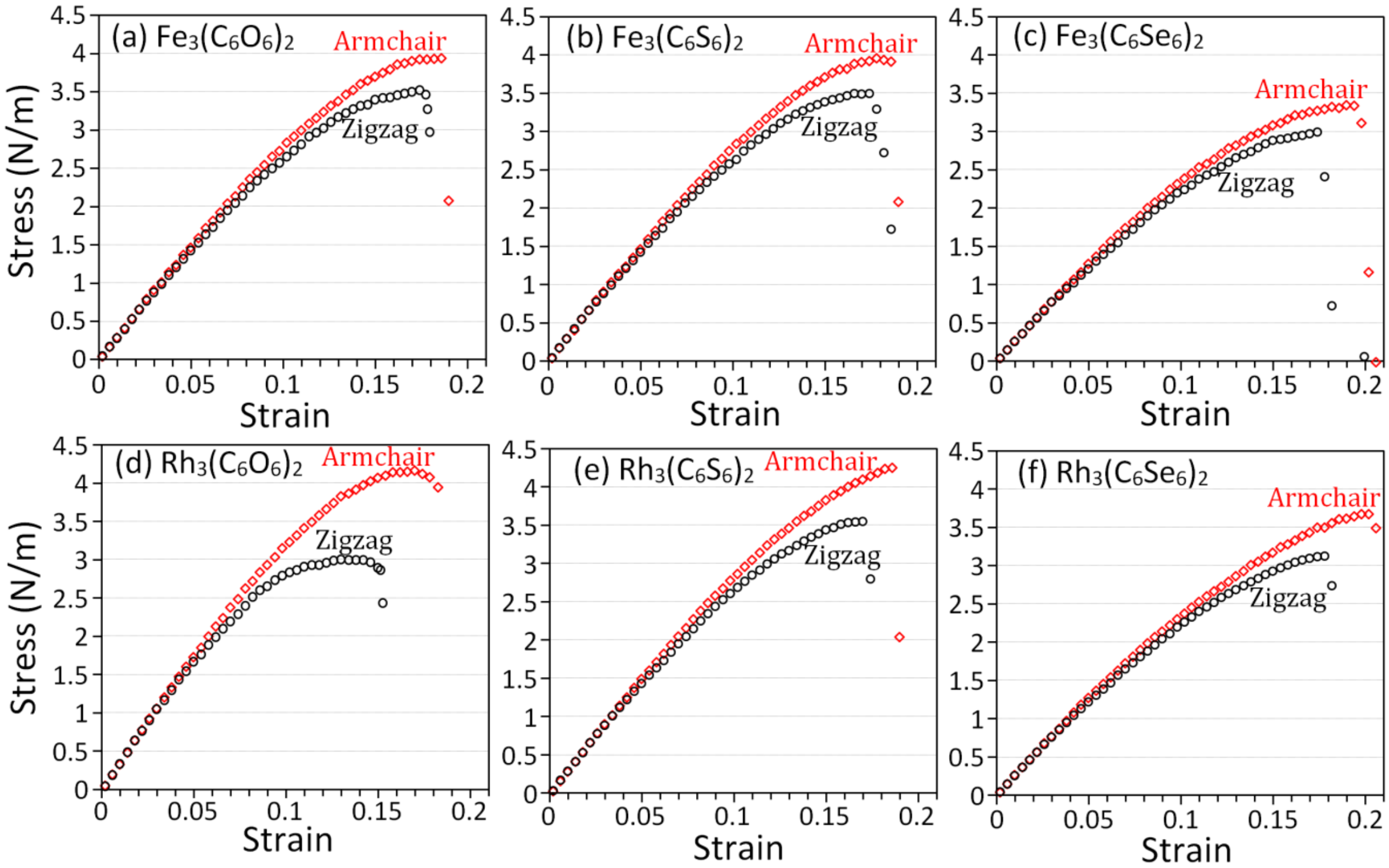

Fig. 2, Uniaxial stress-strain responses of 6 different $M_3(C_6X_6)_2$ monolayers.

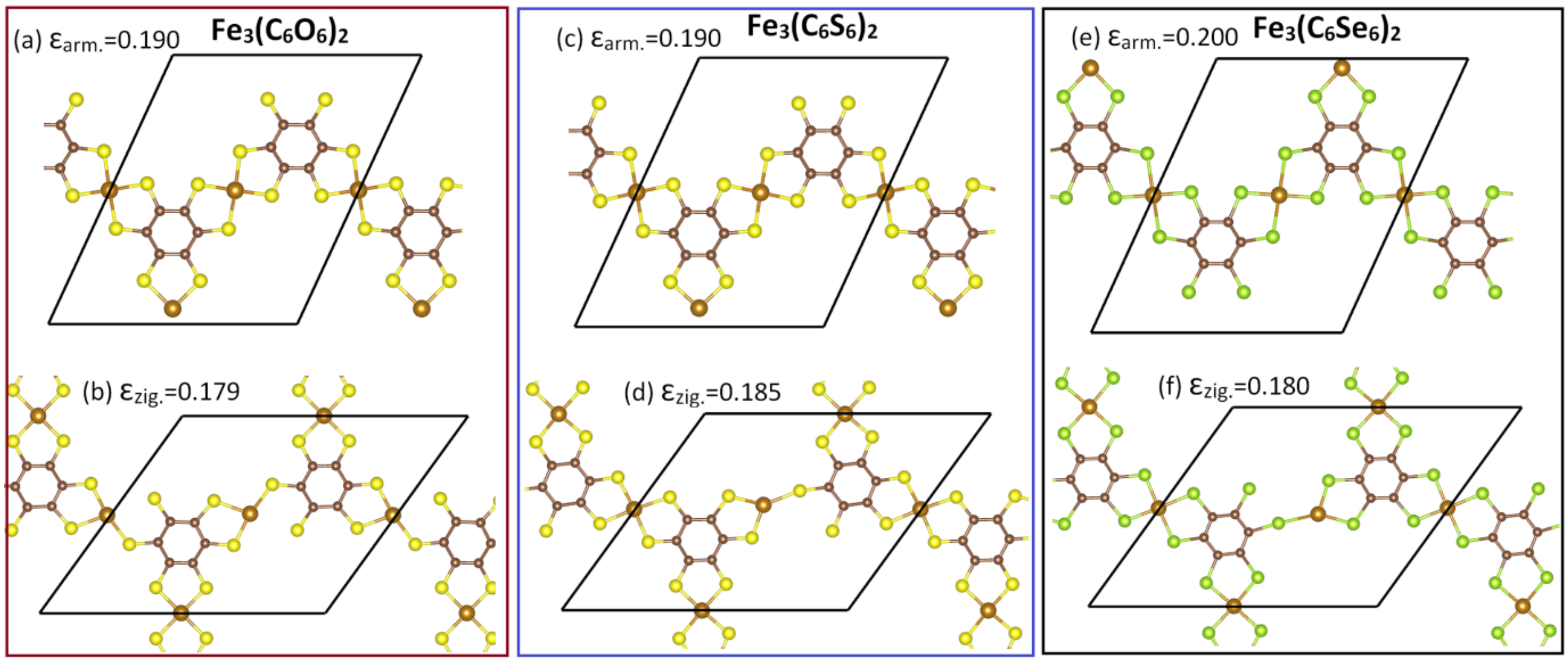

Fig. 3, Top views of the single-layer $Fe_3(C_6X_6)_2$(X=O, S, Se) at strain levels shortly after the ultimate tensile strength point. $\varepsilon_{arm.}$ and $\varepsilon_{zig.}$ depict the strain levels for the uniaxial loading along the armchair and zigzag directions, respectively.

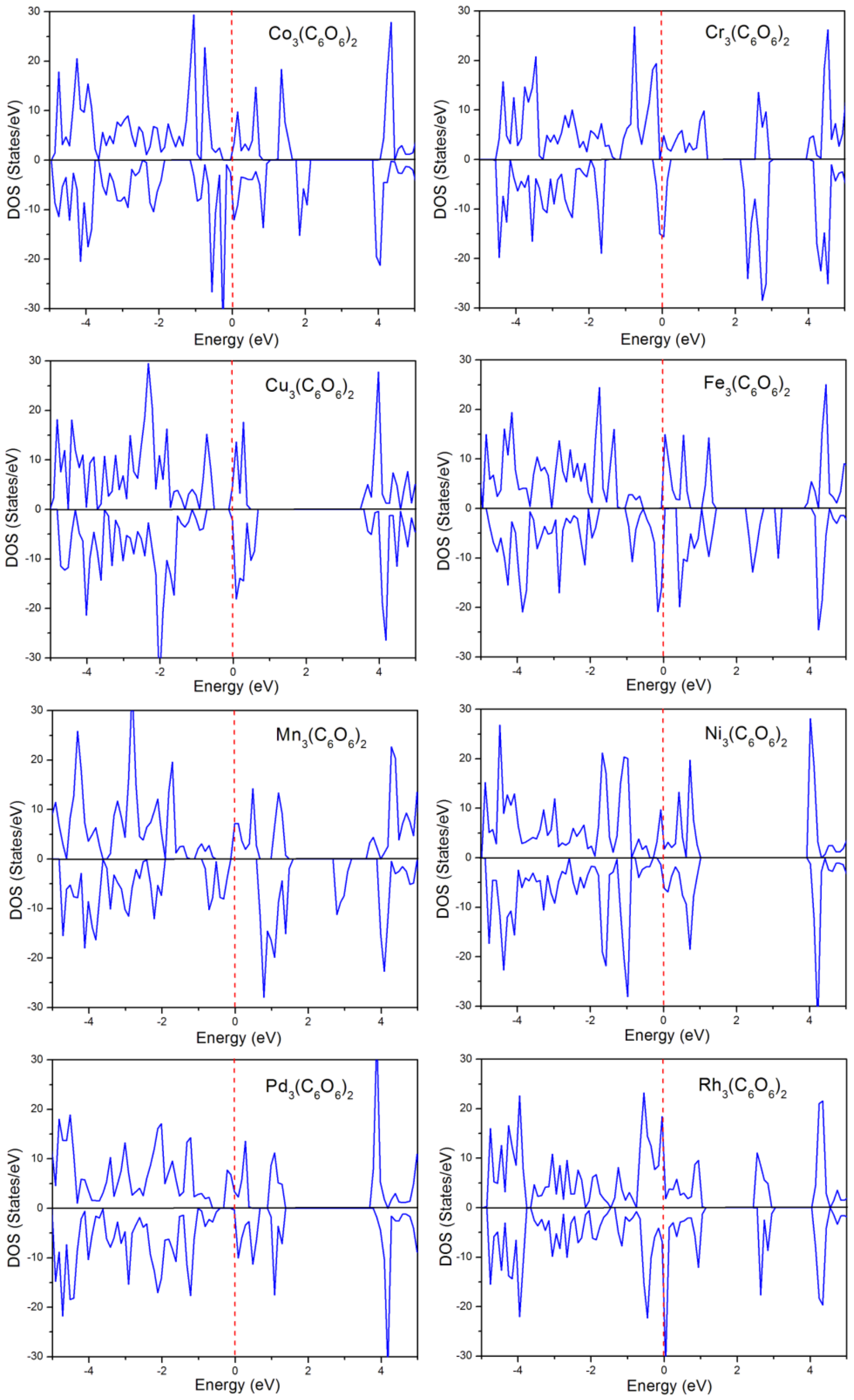

Fig. 4, Total electronic density of states of single-layer $M_3(C_6O_6)_2$. The Fermi energy is aligned to zero.

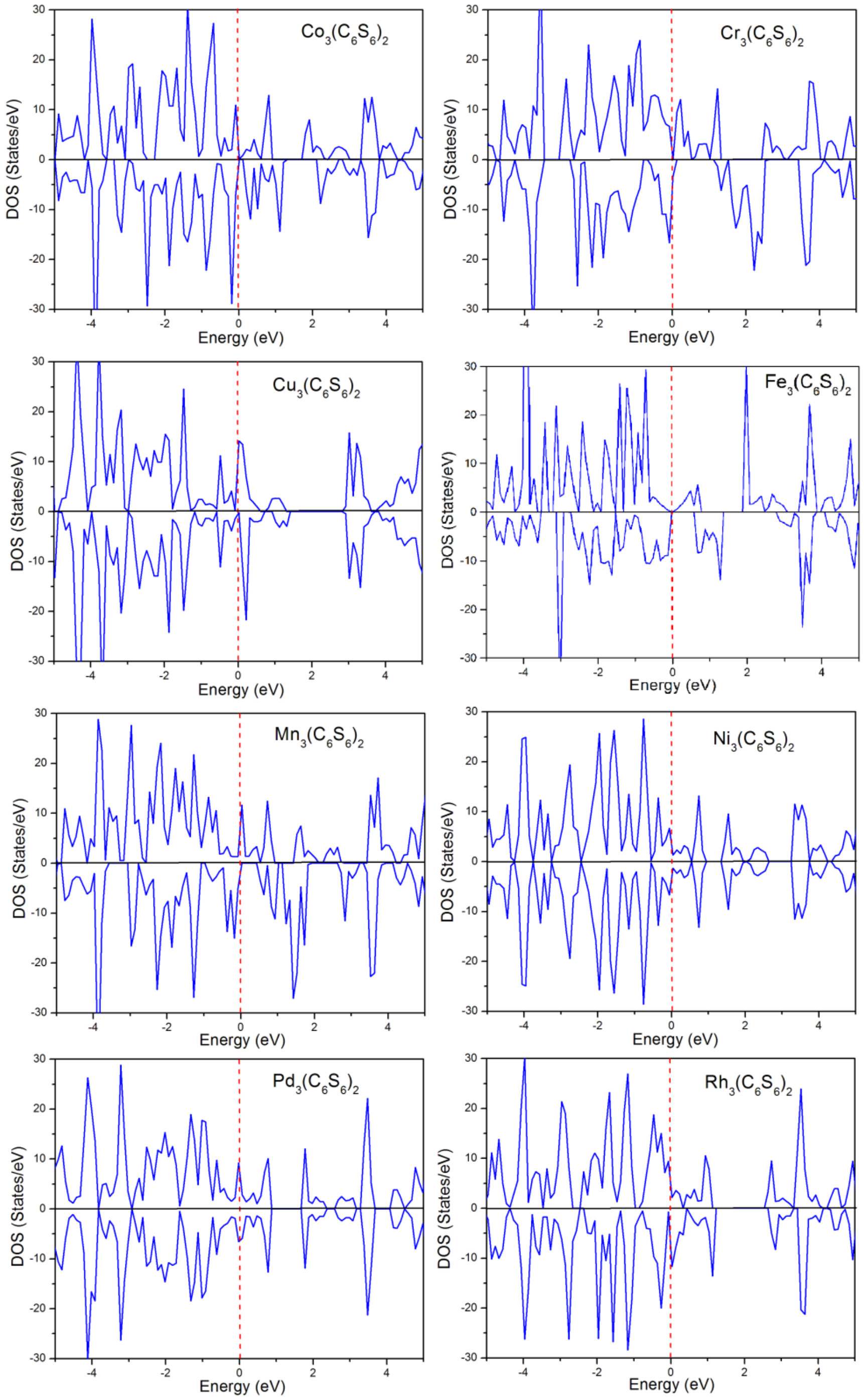

Fig. 5, Total electronic density of states of single-layer $M_3(C_6S_6)_2$. The Fermi energy is aligned to zero.

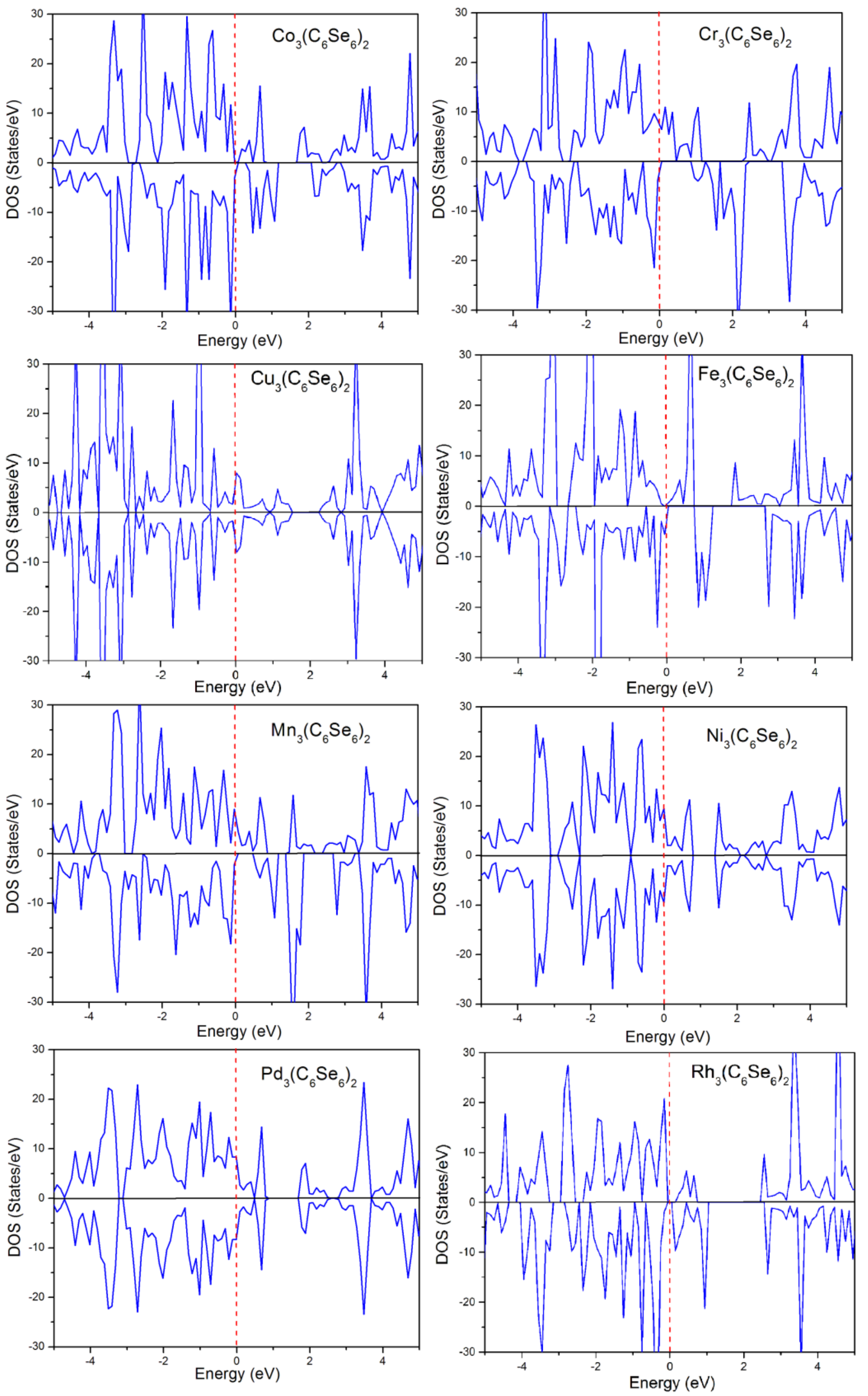

Fig. 6, Total electronic density of states of single-layer $M_3(C_6Se_6)_2$. The Fermi energy is aligned to zero.

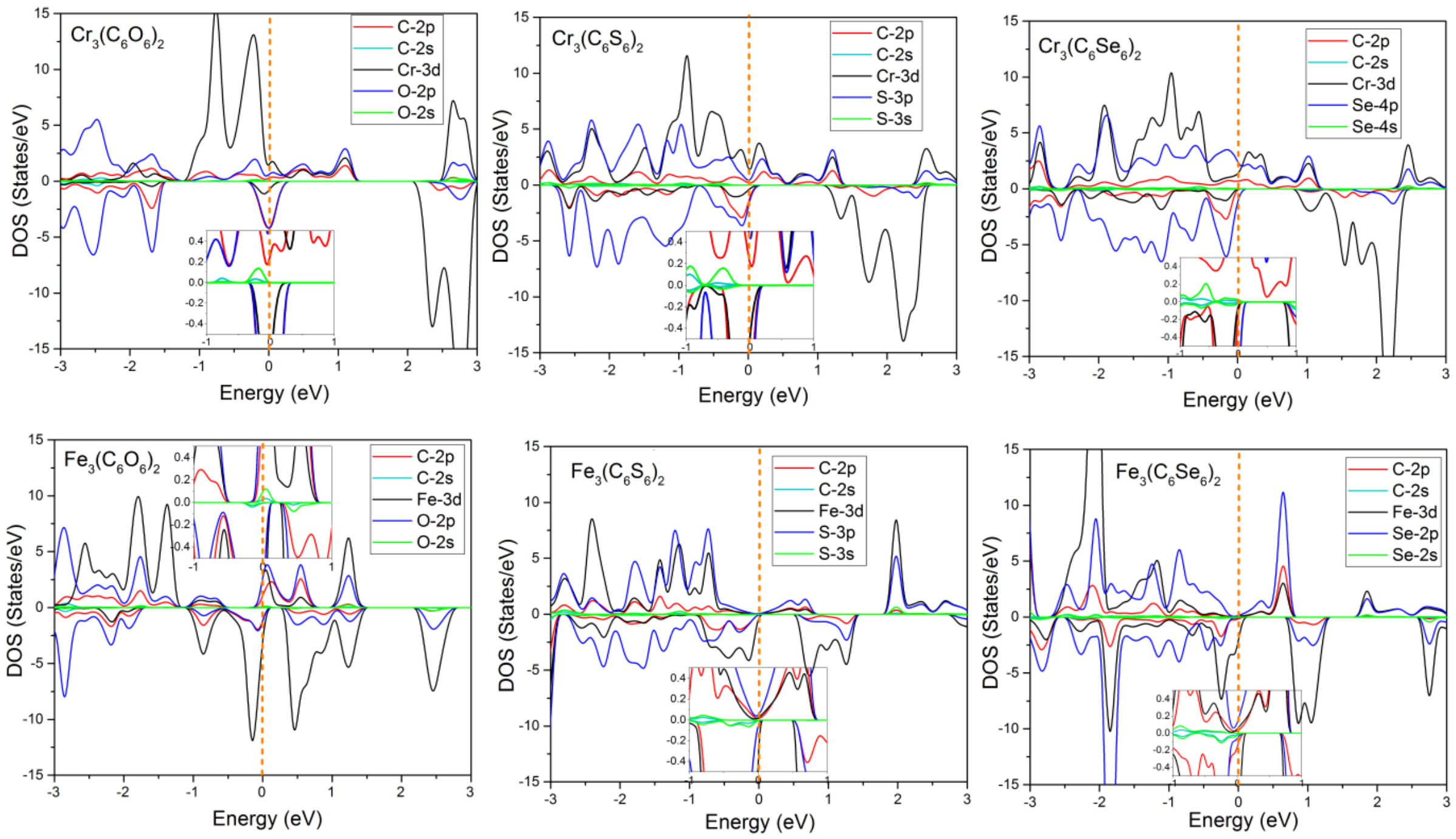

Fig. 7, Partial DOS for $Cr_3(C_6X_6)_2$ and $Fe_3(C_6X_6)_2$ monolayers. The Fermi energy is aligned to zero.

# Supporting Information

# Theoretical realization of two-dimensional $M_3(C_6X_6)_2$ (M= Co, Cr, Cu, Fe, Mn, Ni, Pd, Rh and X= O, S, Se) metal-organic frameworks

Bohayra Mortazavi[*b,c], Masoud Shahrokhi[d], Tanveer Hussain[e,f], Xiaoying Zhuang[c] and Timon Rabczuk[a]

[a]*Institute of Research & Development, Duy Tan University, Quang Trung, Da Nang, Vietnam.*

[b]*Institute of Structural Mechanics, Bauhaus-Universität Weimar, Marienstr. 15, D-99423 Weimar, Germany.*

[c]*Cluster of Excellence PhoenixD (Photonics, Optics, and Engineering–Innovation Across Disciplines), Gottfried Wilhelm Leibniz Universität Hannover, Hannover, Germany.*

[d]*Department of Physics, Faculty of Science, Razi University, Kermanshah, Iran.*

[e]*School of Molecular Sciences, The University of Western Australia, Perth, WA 6009, Australia. Centre for* [f]*Theoretical and Computational Molecular Science, Australian Institute for Bioengineering and Nanotechnology, The University of Queensland, Brisbane, Queensland 4072, Australia*

*E-mail: bohayra.mortazavi@gmail.com

## 1. Atomic structures of constructed monolayers unit-cells in VASP POSCAR

```
Co3C12O12
   1.00000000000000
```

```
    13.0463505106483897    0.0000000000000000    0.0000000000000000
     6.5231752553731308   11.2984709689281093    0.0000000000000000
     0.0000000000000000    0.0000000000000000   16.0000000000000000
   C    O    Co
    12    12     3
Direct
  0.3303438798406262  0.2246438563500703  0.5000000000000000
  0.4409431354388147  0.2245662814712048  0.5000000000000000
  0.4410067161713869  0.3340193927224249  0.5000000000000000
  0.3304203429029258  0.4445360808383256  0.5000000000000000
  0.2209922229510966  0.4445280193888905  0.5000000000000000
  0.2209539828946490  0.3340883836241773  0.5000000000000000
  0.7739441237748963  0.5575726690300726  0.5000000000000000
  0.7739225641100447  0.6681317365867514  0.5000000000000000
  0.6644090384276851  0.7775424831396407  0.5000000000000000
  0.5539150840283753  0.7775039238223513  0.5000000000000000
  0.5539402628488475  0.6681194750895614  0.5000000000000000
  0.6644452792414555  0.5576191840008491  0.5000000000000000
  0.3408212748246342  0.1198379182102585  0.5000000000000000
  0.5351152552424949  0.1196278347848789  0.5000000000000000
  0.5352237400179050  0.3446945705819005  0.5000000000000000
  0.3409314118633304  0.5388455814021498  0.5000000000000000
  0.1161006016653374  0.5387686325410996  0.5000000000000000
  0.1160351543650080  0.3447671226483919  0.5000000000000000
  0.8788074428532511  0.4633247466279116  0.5000000000000000
  0.8787959440287025  0.6575318872174876  0.5000000000000000
  0.6536401358590140  0.8824730915397510  0.5000000000000000
  0.4595907882776160  0.8822997447566010  0.5000000000000000
  0.4596385663280600  0.6575700263512942  0.5000000000000000
  0.6537365596525575  0.4634186606329678  0.5000000000000000
  0.4972261192743943  0.0010753759336026  0.5000000000000000
  0.4973458428282740  0.5011687616720053  0.5000000000000000
  0.9974253548756309  0.5011148249538309  0.5000000000000000
```

**Co3C12S12**

```
   1.00000000000000
    14.6786061158087104    0.0000000000000000    0.0000000000000000
     7.3393030579594116   12.7120457884703093    0.0000000000000000
     0.0000000000000000    0.0000000000000000   16.0000000000000000
   C    S    Co
    12    12     3
Direct
```

```
  0.3309431049759084  0.2372546576634491  0.5000000000000000
  0.4277197566775399  0.2372083191777747  0.5000000000000000
  0.4277349044218468  0.3346094223104927  0.5000000000000000
  0.3309637820675277  0.4313913286984388  0.5000000000000000
  0.2336134471720968  0.4313760171456167  0.5000000000000000
  0.2336167191248961  0.3346494376131872  0.5000000000000000
  0.7612648280906740  0.5707847590517758  0.5000000000000000
  0.7612624334174143  0.6675696329893483  0.5000000000000000
  0.6638441374341468  0.7649431404498657  0.5000000000000000
  0.5671143193122958  0.7648913255635819  0.5000000000000000
  0.5671361411783522  0.6675481588262002  0.5000000000000000
  0.6638887380439442  0.5707992016511056  0.5000000000000000
  0.3342614274072417  0.1185187820272802  0.5000000000000000
  0.5430069705168705  0.1183826360982323  0.5000000000000000
  0.5430434933823634  0.3380814633040998  0.5000000000000000
  0.3343487422366493  0.5467565531907042  0.5000000000000000
  0.1148057096956734  0.5467347507592493  0.5000000000000000
  0.1148226408009876  0.3380538592557798  0.5000000000000000
  0.8800182565559282  0.4554174921019225  0.5000000000000000
  0.8800357189833505  0.6641051235153625  0.5000000000000000
  0.6603044694035773  0.8837582283780847  0.5000000000000000
  0.4517010343640990  0.8836675707460344  0.5000000000000000
  0.4517504236971561  0.6641236866421281  0.5000000000000000
  0.6603874131170784  0.4554904123701249  0.5000000000000000
  0.4972715495478681  0.0010855908537835  0.5000000000000000
  0.4973749606559679  0.5011183472357672  0.5000000000000000
  0.9974178889842733  0.5010704621159050  0.5000000000000000
```

**Co3C12Se12**

```
1.00000000000000
    15.3832960545709092    0.0000000000000000    0.0000000000000000
     7.6916480272854590   13.3223251771956299    0.0000000000000000
     0.0000000000000000    0.0000000000000000   16.0000000000000000
   C    Se   Co
    12    12     3
Direct
  0.3310572895220005  0.2422307537286130  0.5000000000000000
  0.4226742827869359  0.2422016144998906  0.5000000000000000
  0.4226445538750454  0.3347348571149666  0.5000000000000000
```

```
  0.3310323294352742  0.4263322714072260  0.5000000000000000
  0.2385285502041725  0.4263464145088633  0.5000000000000000
  0.2385463324316532  0.3347221285290018  0.5000000000000000
  0.7562580844506002  0.5758412709578522  0.5000000000000000
  0.7562857053888195  0.6674448348956759  0.5000000000000000
  0.6637793457625776  0.7599664123751140  0.5000000000000000
  0.5721493826858506  0.7600121896277727  0.5000000000000000
  0.5721249157591544  0.6674947299645382  0.5000000000000000
  0.6637252864477290  0.5758797029899583  0.5000000000000000
  0.3322981349850949  0.1186044622952664  0.5000000000000000
  0.5450114031412170  0.1185539292890212  0.5000000000000000
  0.5449847794752358  0.3360181652041234  0.5000000000000000
  0.3322908394316144  0.5486910608539475  0.5000000000000000
  0.1148570732639556  0.5486451046643310  0.5000000000000000
  0.1149287879259902  0.3359653718724687  0.5000000000000000
  0.8798610565736382  0.4534835206460457  0.5000000000000000
  0.8799246562685994  0.6661071629420618  0.5000000000000000
  0.6624878436566988  0.8835946661538259  0.5000000000000000
  0.4497934423334087  0.8836031123237476  0.5000000000000000
  0.4497842929409614  0.6661897937209176  0.5000000000000000
  0.6624500766482981  0.4535229051292262  0.5000000000000000
  0.4973918937788255  0.0010847986150040  0.5000000000000000
  0.4973815805557393  0.5010983643769862  0.5000000000000000
  0.9974010915366165  0.5010207610487925  0.5000000000000000
```

**Cr3C12O12**

```
   1.00000000000000
    13.4576135277310200    0.0000000000000000    0.0000000000000000
     6.7288067639159870   11.6546351893596594    0.0000000000000000
     0.0000000000000000    0.0000000000000000   16.0000000000000000
   C     O     Cr
    12    12     3
Direct
  0.3301992688595876  0.2277843444964305  0.5000000000000000
  0.4378193909961797  0.2277641865984882  0.5000000000000000
  0.4378286697778200  0.3339562941298119  0.5000000000000000
  0.3302626428763678  0.4415317787722519  0.5000000000000000
  0.2240103866811314  0.4415976989879955  0.5000000000000000
  0.2239803942943840  0.3339913878749596  0.5000000000000000
  0.7708153669469269  0.5606211191531569  0.5000000000000000
```

```
  0.7707945917923595  0.6681958813259287  0.5000000000000000
  0.6645726642926493  0.7743998164944177  0.5000000000000000
  0.5569191850064609  0.7744098109013207  0.5000000000000000
  0.5569397347651691  0.6681976478226084  0.5000000000000000
  0.6646040712389084  0.5605909999695798  0.5000000000000000
  0.3393943426038789  0.1265703041760560  0.5000000000000000
  0.5298302225601093  0.1265571045527743  0.5000000000000000
  0.5298541122417449  0.3431537415669900  0.5000000000000000
  0.3394787956956584  0.5335232688559420  0.5000000000000000
  0.1228394957288600  0.5336569467923890  0.5000000000000000
  0.1227576673575618  0.3431903431771732  0.5000000000000000
  0.8720311107485721  0.4685881667596865  0.5000000000000000
  0.8720213812542738  0.6590278056350020  0.5000000000000000
  0.6553326500772484  0.8756200696531532  0.5000000000000000
  0.4648959102570985  0.8756542334597965  0.5000000000000000
  0.4649584456591682  0.6588951754520603  0.5000000000000000
  0.6554304831968258  0.4685802613971606  0.5000000000000000
  0.4973223114672774  0.0011230943821538  0.5000000000000000
  0.4973975691742410  0.5010627099418996  0.5000000000000000
  0.9973792321797527  0.5011548038550117  0.5000000000000000
```

**Cr3C12S12**

```
   1.00000000000000
     15.1070571207800803    0.0000000000000000    0.0000000000000000
      7.5535285604466802   13.0830952430534708    0.0000000000000000
      0.0000000000000000    0.0000000000000000   16.0000000000000000
   C    S    Cr
    12    12     3
Direct
  0.3306410957800166  0.2399118135999459  0.5000000000000000
  0.4252827604099345  0.2399187298203212  0.5000000000000000
  0.4252755102350129  0.3343183702282181  0.5000000000000000
  0.3306545342210623  0.4289659071778463  0.5000000000000000
  0.2362900935607956  0.4289719313817062  0.5000000000000000
  0.2362905901338308  0.3342974505459291  0.5000000000000000
  0.7584834387015601  0.5732559619529394  0.5000000000000000
  0.7584644816699539  0.6678951018542300  0.5000000000000000
  0.6640876706713721  0.7622691776843353  0.5000000000000000
  0.5694234252842805  0.7622960626699341  0.5000000000000000
  0.5694312326569398  0.6678691708524954  0.5000000000000000
```

```
  0.6641260101692683  0.5732370816971160  0.5000000000000000
  0.3300211160345715  0.1259242715829258  0.5000000000000000
  0.5399203726150631  0.1259998655065602  0.5000000000000000
  0.5399148247502485  0.3336047601832491  0.5000000000000000
  0.3301046752393096  0.5435433632030502  0.5000000000000000
  0.1223367794012075  0.5435938466565550  0.5000000000000000
  0.1222995270347491  0.3336895755026016  0.5000000000000000
  0.8724918131596284  0.4586406586236631  0.5000000000000000
  0.8724627886435764  0.6685216528553866  0.5000000000000000
  0.6647827928080717  0.8762385504416204  0.5000000000000000
  0.4548322845568566  0.8762727263854607  0.5000000000000000
  0.4548756325345863  0.6684069898858828  0.5000000000000000
  0.6648256834460166  0.4585821947014564  0.5000000000000000
  0.4974453338732918  0.0010932228842768  0.5000000000000000
  0.4975166299631729  0.5009576095143313  0.5000000000000000
  0.9974278579791287  0.5011146249228814  0.5000000000000000
```

**Cr3C12Se12**

```
   1.00000000000000
     15.7796932825064502    0.0000000000000000    0.0000000000000000
      7.8898466412532384   13.6656152465353298    0.0000000000000000
      0.0000000000000000    0.0000000000000000   16.0000000000000000
   C    Se   Cr
    12    12     3
Direct
  0.3307269575898175  0.2447087418807214  0.5000000000000000
  0.4204610430696718  0.2447071335185242  0.5000000000000000
  0.4204252210120387  0.3344702660295837  0.5000000000000000
  0.3306881488930884  0.4241691689296587  0.5000000000000000
  0.2409349674792765  0.4241596601719735  0.5000000000000000
  0.2409506099863350  0.3344459902902415  0.5000000000000000
  0.7538115519647235  0.5780350478298288  0.5000000000000000
  0.7537770474243928  0.6677773891001166  0.5000000000000000
  0.6639869800553129  0.7574976272662823  0.5000000000000000
  0.5742640570971105  0.7574783838255001  0.5000000000000000
  0.5743139949819920  0.6677187253796006  0.5000000000000000
  0.6640617752133835  0.5780330504116193  0.5000000000000000
  0.3287377527925841  0.1249463747965649  0.5000000000000000
  0.5422501895936733  0.1249618045752641  0.5000000000000000
  0.5422548199677593  0.3323597663919671  0.5000000000000000
```

```
  0.3287376656004568  0.5459149843064637  0.5000000000000000
  0.1212471359723750  0.5459726266986280  0.5000000000000000
  0.1212416326495429  0.3323679314894576  0.5000000000000000
  0.8735476870454661  0.4562467572379134  0.5000000000000000
  0.8735049052560271  0.6698605380631051  0.5000000000000000
  0.6661161449061126  0.8771891398568599  0.5000000000000000
  0.4525175137021193  0.8772479967788698  0.5000000000000000
  0.4525561984130704  0.6697458704489136  0.5000000000000000
  0.6662053274245139  0.4561885245262687  0.5000000000000000
  0.4974607259866204  0.0010651062347122  0.5000000000000000
  0.4975112585157007  0.5010138094868779  0.5000000000000000
  0.9974176429403059  0.5011082567894505  0.5000000000000000
```

**Cu3C12O12**

```
   1.00000000000000
    13.4537722672539708    0.0000000000000000    0.0000000000000000
     6.7268861336774703   11.6513085602039705    0.0000000000000000
     0.0000000000000000    0.0000000000000000   16.0000000000000000
   C    O    Cu
    12    12     3
Direct
  0.3303166122592387  0.2261256155899218  0.5000000000000000
  0.4393803164567913  0.2260906158136251  0.5000000000000000
  0.4394045073030171  0.3341029911938165  0.5000000000000000
  0.3303580214155772  0.4431654943524208  0.5000000000000000
  0.2222747081826313  0.4432210618171410  0.5000000000000000
  0.2222681163316480  0.3341866353630678  0.5000000000000000
  0.7725020449071636  0.5590525748266586  0.5000000000000000
  0.7724659990316809  0.6680784184558490  0.5000000000000000
  0.6644376582133305  0.7760949275023634  0.5000000000000000
  0.5554189384705808  0.7760449150889599  0.5000000000000000
  0.5554658743205607  0.6680011812049713  0.5000000000000000
  0.6644999857088365  0.5589772925670573  0.5000000000000000
  0.3350291499934883  0.1291483141292673  0.5000000000000000
  0.5315851211175300  0.1290346297280536  0.5000000000000000
  0.5316713821135011  0.3388478445860486  0.5000000000000000
  0.3351105354604869  0.5354128576603827  0.5000000000000000
  0.1252861522563222  0.5355199780632773  0.5000000000000000
  0.1252714829791088  0.3388878267412565  0.5000000000000000
  0.8695124243551930  0.4667678089799594  0.5000000000000000
```

```
  0.8694681212374675  0.6633865247697415  0.5000000000000000
  0.6596854817383786  0.8730870948882057  0.5000000000000000
  0.4630926867205645  0.8729833881467357  0.5000000000000000
  0.4631906644587005  0.6632792015047713  0.5000000000000000
  0.6597805211983427  0.4667117815282182  0.5000000000000000
  0.4973040589605944  0.0010604296293621  0.5000000000000000
  0.4974379725671696  0.5010559331119424  0.5000000000000000
  0.9973781147577299  0.5011518086142530  0.5000000000000000
```

**Cu3C12S12**

```
   1.00000000000000
    14.7103575015763592    0.0000000000000000    0.0000000000000000
     7.3551787508433319   12.7395432951504706    0.0000000000000000
     0.0000000000000000    0.0000000000000000   16.0000000000000000
   C    S    Cu
    12    12     3
Direct
  0.3310978299994259  0.2375838353595003  0.5000000000000000
  0.4271972330621709  0.2375470219926044  0.5000000000000000
  0.4272336653569594  0.3347740696578143  0.5000000000000000
  0.3311520237633380  0.4308685477530094  0.5000000000000000
  0.2339354198953030  0.4308834592980347  0.5000000000000000
  0.2339256981984442  0.3348115657768301  0.5000000000000000
  0.7609277817226601  0.5712602566359147  0.5000000000000000
  0.7608962919771135  0.6673975612682762  0.5000000000000000
  0.6636422563880271  0.7646122274869899  0.5000000000000000
  0.5675779105314718  0.7645924399736390  0.5000000000000000
  0.5676211452501423  0.6673644711175648  0.5000000000000000
  0.6637023993571056  0.5712757197858045  0.5000000000000000
  0.3294725699835719  0.1217201677558018  0.5000000000000000
  0.5445985272276574  0.1215772154484232  0.5000000000000000
  0.5446623402146775  0.3332842044402614  0.5000000000000000
  0.3296344677015934  0.5483103021250977  0.5000000000000000
  0.1179843484923827  0.5483310001354492  0.5000000000000000
  0.1179457110013767  0.3333516557481301  0.5000000000000000
  0.8768396428101732  0.4538744217194535  0.5000000000000000
  0.8767736263171633  0.6689577199167426  0.5000000000000000
  0.6650911675669420  0.8805654477681146  0.5000000000000000
  0.4501041055664045  0.8804900440570130  0.5000000000000000
  0.4501946456735423  0.6688314268788625  0.5000000000000000
```

```
  0.6651991668078033  0.4538557132452082  0.5000000000000000
  0.4972722359993215  0.0010948318079553  0.5000000000000000
  0.4974222925235487  0.5010627175532102  0.5000000000000000
  0.9973820094128489  0.5011504922702933  0.5000000000000000
```

**Cu3C12Se12**

```
   1.00000000000000
     15.3786615884746993    0.0000000000000000    0.0000000000000000
      7.6893307942373639   13.3183116118729092    0.0000000000000000
      0.0000000000000000    0.0000000000000000   16.0000000000000000
   C    Se   Cu
    12    12     3
Direct
  0.3310792747732023  0.2423354102845039  0.5000000000000000
  0.4224511191285671  0.2423479017104242  0.5000000000000000
  0.4224264558527366  0.3348395852283090  0.5000000000000000
  0.3310428439359683  0.4261947416013925  0.5000000000000000
  0.2385781673972076  0.4261801823401186  0.5000000000000000
  0.2385800159426807  0.3347856397576265  0.5000000000000000
  0.7561818741707498  0.5759776899795028  0.5000000000000000
  0.7562191651023014  0.6673491730293080  0.5000000000000000
  0.6637469183589673  0.7598479832279637  0.5000000000000000
  0.5723471890929233  0.7598938116260925  0.5000000000000000
  0.5722982481854686  0.6674113717973569  0.5000000000000000
  0.6636694967284242  0.5760265114453205  0.5000000000000000
  0.3278156854947341  0.1213508680903672  0.5000000000000000
  0.5467164719759054  0.1214075297891242  0.5000000000000000
  0.5466953556479410  0.3314964027068186  0.5000000000000000
  0.3277287299437646  0.5504609722239238  0.5000000000000000
  0.1176237829297690  0.5504195908036920  0.5000000000000000
  0.1176645762506254  0.3314155142198061  0.5000000000000000
  0.8770995140332332  0.4517152105324982  0.5000000000000000
  0.8771619900656447  0.6706413379485525  0.5000000000000000
  0.6670528680600043  0.8808101922108378  0.5000000000000000
  0.4480882716452044  0.8808288328756859  0.5000000000000000
  0.4480400937387756  0.6707117856131930  0.5000000000000000
  0.6669784599943185  0.4517649710527181  0.5000000000000000
  0.4974362341132732  0.0010908342373881  0.5000000000000000
  0.4973665861357830  0.5010989970512725  0.5000000000000000
  0.9973971241031805  0.5010254955922022  0.5000000000000000
```

**Fe3C12O12**

```
   1.00000000000000
    13.1625885132606104    0.0000000000000000    0.0000000000000000
     6.5812942566796711   11.3991360320757007    0.0000000000000000
     0.0000000000000000    0.0000000000000000   16.0000000000000000
   C    O    Fe
    12    12     3
Direct
  0.3304513440396804  0.2262209040576266  0.5000000000000000
  0.4393473327256459  0.2261049729735944  0.5000000000000000
  0.4393434445953162  0.3340867068686961  0.5000000000000000
  0.3304537640502829  0.4429533763758542  0.5000000000000000
  0.2224438957322334  0.4430488213350259  0.5000000000000000
  0.2224790377887444  0.3342167194960268  0.5000000000000000
  0.7723474654478650  0.5591493396576794  0.5000000000000000
  0.7722922989669883  0.6680499110316731  0.5000000000000000
  0.6642802864881503  0.7760158636700325  0.5000000000000000
  0.5554306196427845  0.7760340548523175  0.5000000000000000
  0.5555159077138043  0.6680328080151199  0.5000000000000000
  0.6643639713617731  0.5591828026208603  0.5000000000000000
  0.3422637454377835  0.1207662290382672  0.5000000000000000
  0.5328822595495311  0.1205493574606038  0.5000000000000000
  0.5329101997694536  0.3460421024838354  0.5000000000000000
  0.3423592138708500  0.5365354217430536  0.5000000000000000
  0.1169362519148862  0.5366721993993266  0.5000000000000000
  0.1169901417335808  0.3460874500528845  0.5000000000000000
  0.8778762672285154  0.4655312638334337  0.5000000000000000
  0.8777578035151521  0.6562389217960458  0.5000000000000000
  0.6523712388548475  0.8815160126375545  0.5000000000000000
  0.4617836986058279  0.8814899320305543  0.5000000000000000
  0.4619359326187791  0.6561383774319367  0.5000000000000000
  0.6524492433285189  0.4655947248165688  0.5000000000000000
  0.4972801010103893  0.0010807739959162  0.5000000000000000
  0.4973959081016588  0.5010916723663428  0.5000000000000000
  0.9973846681067613  0.5011421748142766  0.5000000000000000
```

**Fe3C12S12**

```
   1.00000000000000
    14.8060775256745796    0.0000000000000000    0.0000000000000000
     7.4030387628928356   12.8224392676706493    0.0000000000000000
     0.0000000000000000    0.0000000000000000   16.0000000000000000
   C    S    Fe
    12    12     3
Direct
  0.3309077730416874  0.2381261858182384  0.5000000000000000
  0.4267758127175796  0.2381087649582128  0.5000000000000000
  0.4268014994837088  0.3346532220843059  0.5000000000000000
  0.3309483182540447  0.4305269274714121  0.5000000000000000
  0.2343993622786158  0.4305518102860262  0.5000000000000000
  0.2344092573053302  0.3346769463737402  0.5000000000000000
  0.7603771370914316  0.5716818585873327  0.5000000000000000
  0.7603850029090040  0.6675598516913439  0.5000000000000000
  0.6638423431271860  0.7640691704941389  0.5000000000000000
  0.5679760748186382  0.7640622901613341  0.5000000000000000
  0.5679973817465438  0.6675263609428228  0.5000000000000000
  0.6638622264889165  0.5716770446023176  0.5000000000000000
  0.3331426884198336  0.1205456221424441  0.5000000000000000
  0.5420651199664235  0.1205034901664064  0.5000000000000000
  0.5421252992788581  0.3369058234777356  0.5000000000000000
  0.3332305769788206  0.5458464144198274  0.5000000000000000
  0.1167978572842330  0.5458695303411929  0.5000000000000000
  0.1168191490551109  0.3369337812221360  0.5000000000000000
  0.8779343833163225  0.4563599711205342  0.5000000000000000
  0.8779479084875916  0.6652951025068035  0.5000000000000000
  0.6615338110572679  0.8816638076776044  0.5000000000000000
  0.4526132454288501  0.8816016325538385  0.5000000000000000
  0.4526756398119645  0.6652405056533865  0.5000000000000000
  0.6616166973440656  0.4563432595769825  0.5000000000000000
  0.4973237108717399  0.0010750261840826  0.5000000000000000
  0.4974323168253463  0.5010655199174820  0.5000000000000000
  0.9973854488107889  0.5011029744234961  0.5000000000000000
```

**Fe3C12Se12**

```
   1.00000000000000
    15.5178759132559101    0.0000000000000000    0.0000000000000000
     7.7589379566279648   13.4388747536908895    0.0000000000000000
     0.0000000000000000    0.0000000000000000   16.0000000000000000
   C    Se   Fe
    12    12     3
Direct
  0.3309968621599921  0.2430062342399708  0.5000000000000000
  0.4218460371836557  0.2430051378772404  0.5000000000000000
  0.4218546566958068  0.3347478976729050  0.5000000000000000
  0.3310094147760623  0.4255676215368212  0.5000000000000000
  0.2392672331804775  0.4255710803297461  0.5000000000000000
  0.2392607415452730  0.3347328077296439  0.5000000000000000
  0.7554768788429982  0.5766157333825319  0.5000000000000000
  0.7554822715294520  0.6674678779491359  0.5000000000000000
  0.6637453920778995  0.7592008980446749  0.5000000000000000
  0.5728988133130315  0.7592081959844563  0.5000000000000000
  0.5729116410101938  0.6674508635699254  0.5000000000000000
  0.6637497864050275  0.5766147618591191  0.5000000000000000
  0.3311640505917595  0.1206348334746110  0.5000000000000000
  0.5440503241814696  0.1206447828246482  0.5000000000000000
  0.5440810430148915  0.3348402428799133  0.5000000000000000
  0.3311885824847067  0.5477534429563704  0.5000000000000000
  0.1169495759653358  0.5477887563542225  0.5000000000000000
  0.1169164190948138  0.3348934524890979  0.5000000000000000
  0.8778340192729317  0.4544240961262034  0.5000000000000000
  0.8778060684576232  0.6673524563120594  0.5000000000000000
  0.6636035545815631  0.8815536854604815  0.5000000000000000
  0.4506900623143437  0.8815294494145647  0.5000000000000000
  0.4507247123842291  0.6673008451571931  0.5000000000000000
  0.6636433978336171  0.4543909446743939  0.5000000000000000
  0.4973780444896576  0.0010872003320586  0.5000000000000000
  0.4974174143760592  0.5010643175034986  0.5000000000000000
  0.9973790444370181  0.5011252787196980  0.5000000000000000
```

**Mn3C12O12**

```
   1.00000000000000
    13.2507894106378998    0.0000000000000000    0.0000000000000000
     6.6253947053686417   11.4755202498412192    0.0000000000000000
     0.0000000000000000    0.0000000000000000   16.0000000000000000
   C    O    Mn
    12    12     3
Direct
  0.3302723043093323  0.2271409071232355  0.5000000000000000
  0.4383914351632114  0.2271248843922606  0.5000000000000000
  0.4384025575141095  0.3340301377659856  0.5000000000000000
  0.3303031140222359  0.4421733699433474  0.5000000000000000
  0.2232898393732086  0.4422486668685224  0.5000000000000000
  0.2232818590596466  0.3341085362184018  0.5000000000000000
  0.7713953925404411  0.5600168570732009  0.5000000000000000
  0.7713699530155935  0.6681852025247750  0.5000000000000000
  0.6644398836165735  0.7751595375058713  0.5000000000000000
  0.5563920482507569  0.7750673868744471  0.5000000000000000
  0.5564045763226275  0.6680373802752015  0.5000000000000000
  0.6644874787828812  0.5599245871180827  0.5000000000000000
  0.3410489750303256  0.1222274924479905  0.5000000000000000
  0.5326146778749532  0.1222344744938866  0.5000000000000000
  0.5326746116436780  0.3445961209034962  0.5000000000000000
  0.3410857055587115  0.5363066228607423  0.5000000000000000
  0.1183964621241671  0.5364467407706357  0.5000000000000000
  0.1184085443750149  0.3447935515013540  0.5000000000000000
  0.8762999098692319  0.4658300411956162  0.5000000000000000
  0.8762467014489630  0.6575366457242566  0.5000000000000000
  0.6537609294603541  0.8800761905939325  0.5000000000000000
  0.4622356977130124  0.8799738297105080  0.5000000000000000
  0.4622560936755491  0.6572724801639325  0.5000000000000000
  0.6538232525940018  0.4656722102428077  0.5000000000000000
  0.4973940832926615  0.0011290545074658  0.5000000000000000
  0.4974788862340915  0.5009483168163058  0.5000000000000000
  0.9973315399358711  0.5011673113597439  0.5000000000000000
```

**Mn3C12S12**

```
   1.00000000000000
    14.9826052460489105    0.0000000000000000    0.0000000000000000
     7.4913026230806263   12.9753167579873505    0.0000000000000000
     0.0000000000000000    0.0000000000000000   16.0000000000000000
   C    S    Mn
    12    12     3
Direct
  0.3309062565244039  0.2392284135741215  0.5000000000000000
  0.4256783266515313  0.2392336892562381  0.5000000000000000
  0.4257046876599288  0.3346104246389714  0.5000000000000000
  0.3309644265825635  0.4294459682882832  0.5000000000000000
  0.2356075003260099  0.4294278145154397  0.5000000000000000
  0.2355793187901014  0.3346302886637038  0.5000000000000000
  0.7592450067831030  0.5728056390652837  0.5000000000000000
  0.7592091380401946  0.6675887198952688  0.5000000000000000
  0.6637953859462754  0.7629096508246803  0.5000000000000000
  0.5689863283354840  0.7629527118764331  0.5000000000000000
  0.5690264851055629  0.6675434530557917  0.5000000000000000
  0.6638730948939345  0.5728160421461612  0.5000000000000000
  0.3298672510045064  0.1245688457030039  0.5000000000000000
  0.5413409828860214  0.1245171796000835  0.5000000000000000
  0.5414427247799853  0.3335065576608081  0.5000000000000000
  0.3300165543463244  0.5450908945287734  0.5000000000000000
  0.1209140476912671  0.5451199998042454  0.5000000000000000
  0.1208599922656134  0.3336912859753854  0.5000000000000000
  0.8739535858860752  0.4571398919996696  0.5000000000000000
  0.8738878146393285  0.6686239574794838  0.5000000000000000
  0.6647425000953189  0.8776195917638645  0.5000000000000000
  0.4533147458721504  0.8776097754503406  0.5000000000000000
  0.4534054097891525  0.6684432039023140  0.5000000000000000
  0.6649238648894896  0.4571066374378874  0.5000000000000000
  0.4973093962051514  0.0010821824693832  0.5000000000000000
  0.4975256963603319  0.5009500258042350  0.5000000000000000
  0.9974059904513809  0.5011656915961878  0.5000000000000000
```

**Mn3C12Se12**

```
   1.00000000000000
    15.6773135056107193    0.0000000000000000    0.0000000000000000
```

```
     7.8386567528053606   13.5769517589414992    0.0000000000000000
     0.0000000000000000    0.0000000000000000   16.0000000000000000
   C    Se   Mn
    12    12     3
Direct
  0.3308957413962972  0.2438442700775809  0.5000000000000000
  0.4210705482584984  0.2438138854316350  0.5000000000000000
  0.4210902206848317  0.3346439207380030  0.5000000000000000
  0.3309286834244531  0.4247665662698168  0.5000000000000000
  0.2400505314638437  0.4247986628041360  0.5000000000000000
  0.2400289996233624  0.3346758850047422  0.5000000000000000
  0.7546956309528170  0.5774093858862983  0.5000000000000000
  0.7546864073344324  0.6675589169147713  0.5000000000000000
  0.6638148167166165  0.7583836670804516  0.5000000000000000
  0.5736651257057446  0.7583801797658936  0.5000000000000000
  0.5736900491662382  0.6675181159125287  0.5000000000000000
  0.6638409036080333  0.5773971885593951  0.5000000000000000
  0.3276665701769161  0.1246472603637656  0.5000000000000000
  0.5435352649891954  0.1246261231398407  0.5000000000000000
  0.5435763265995988  0.3313125102246772  0.5000000000000000
  0.3277221837784836  0.5471958291249663  0.5000000000000000
  0.1209422896688395  0.5472927581688936  0.5000000000000000
  0.1208858180183583  0.3313734809812274  0.5000000000000000
  0.8738750806404738  0.4549464107078833  0.5000000000000000
  0.8738554798415024  0.6708546721216777  0.5000000000000000
  0.6670972330419929  0.8775647517752745  0.5000000000000000
  0.4512060764122552  0.8775401666809515  0.5000000000000000
  0.4512649079666460  0.6707496447004289  0.5000000000000000
  0.6671741834319391  0.4549050562491601  0.5000000000000000
  0.4973700398424654  0.0010860492926028  0.5000000000000000
  0.4974610369417576  0.5010140794662448  0.5000000000000000
  0.9973963631156053  0.5011290995332160  0.5000000000000000
```

**Ni3C12O12**

```
   1.00000000000000
    12.9901051162296799    0.0000000000000000    0.0000000000000000
     6.4950525581635663   11.2497610285155201    0.0000000000000000
     0.0000000000000000    0.0000000000000000   16.0000000000000000
   C    O    Ni
    12    12     3
```

```
Direct
  0.3306079108582358  0.2241478684237632  0.5000000000000000
  0.4412145462035483  0.2240501508537222  0.5000000000000000
  0.4412553263118250  0.3342954375658220  0.5000000000000000
  0.3306592680606398  0.4448040549030949  0.5000000000000000
  0.2204120988516865  0.4448217783886186  0.5000000000000000
  0.2203780155206266  0.3343738315823445  0.5000000000000000
  0.7744848976413139  0.5573003728702770  0.5000000000000000
  0.7744768701258821  0.6678606949954599  0.5000000000000000
  0.6641564388169279  0.7780665574838259  0.5000000000000000
  0.5536750026121180  0.7779938559860005  0.5000000000000000
  0.5537065964052336  0.6678064669618587  0.5000000000000000
  0.6641972252687712  0.5573309044397061  0.5000000000000000
  0.3402299727007767  0.1192665838278928  0.5000000000000000
  0.5363207648969119  0.1190750184826422  0.5000000000000000
  0.5363965652576259  0.3441184920943225  0.5000000000000000
  0.3402875315978384  0.5400588441309940  0.5000000000000000
  0.1154804868394313  0.5400858577252921  0.5000000000000000
  0.1154259220183033  0.3440980667827844  0.5000000000000000
  0.8793872662467319  0.4621081418186369  0.5000000000000000
  0.8793917938698984  0.6581419042096773  0.5000000000000000
  0.6542616084623448  0.8830647445520098  0.5000000000000000
  0.4583835299852997  0.8828486626104499  0.5000000000000000
  0.4584515884086137  0.6581541246189555  0.5000000000000000
  0.6543921119690782  0.4621737444263140  0.5000000000000000
  0.4972382564061256  0.0010689676032516  0.5000000000000000
  0.4973477204904748  0.5011602876163024  0.5000000000000000
  0.9974336954394148  0.5011149447811980  0.5000000000000000
```

**Ni3C12S12**

```
   1.00000000000000
    14.6331835023426500    0.0000000000000000    0.0000000000000000
     7.3165917512262144   12.6727086513022993    0.0000000000000000
     0.0000000000000000    0.0000000000000000   16.0000000000000000
   C    S    Ni
    12    12     3
Direct
  0.3310464301701678  0.2366116748029178  0.5000000000000000
  0.4282190787362229  0.2365503873880891  0.5000000000000000
  0.4282572058543250  0.3347699135010400  0.5000000000000000
```

```
  0.3311011743075873  0.4319049857527079  0.5000000000000000
  0.2328716164408978  0.4319462238280218  0.5000000000000000
  0.2328600517165664  0.3348420155357899  0.5000000000000000
  0.7619782178529987  0.5702168613256515  0.5000000000000000
  0.7619669605026276  0.6673925680686160  0.5000000000000000
  0.6637038250887355  0.7655827341108202  0.5000000000000000
  0.5665838441704000  0.7655568684801944  0.5000000000000000
  0.5666041895044505  0.6673737061144359  0.5000000000000000
  0.6637372135018901  0.5702554272266855  0.5000000000000000
  0.3335888442000743  0.1187120699858813  0.5000000000000000
  0.5434886161212379  0.1186098924896593  0.5000000000000000
  0.5435639791815742  0.3373484931312995  0.5000000000000000
  0.3336951850289919  0.5472163139179935  0.5000000000000000
  0.1149693441225348  0.5473049190722979  0.5000000000000000
  0.1149442099220274  0.3374081290497659  0.5000000000000000
  0.8798712815040277  0.4549241895528061  0.5000000000000000
  0.8798312098018258  0.6648507212185145  0.5000000000000000
  0.6610310004460027  0.8835263424293125  0.5000000000000000
  0.4512223962031591  0.8834693514496060  0.5000000000000000
  0.4512773524130542  0.6647642408166454  0.5000000000000000
  0.6611214842410837  0.4549637261864623  0.5000000000000000
  0.4973032057455848  0.0010872225773111  0.5000000000000000
  0.4974146874499798  0.5010681318181802  0.5000000000000000
  0.9974004070376715  0.5011332499045820  0.5000000000000000
```

**Ni3C12Se12**

```
   1.00000000000000
     15.3430810983820507    0.0000000000000000    0.0000000000000000
      7.6715405491910298   13.2874980034864407    0.0000000000000000
      0.0000000000000000    0.0000000000000000   16.0000000000000000
   C    Se   Ni
    12    12     3
Direct
  0.3311970660751982  0.2415329665463872  0.5000000000000000
  0.4232589080565816  0.2415464349101839  0.5000000000000000
  0.4232393810887274  0.3347761729930170  0.5000000000000000
  0.3311450752895766  0.4268178926246264  0.5000000000000000
  0.2379160006702037  0.4268235419362867  0.5000000000000000
  0.2379410195358176  0.3347357959966288  0.5000000000000000
  0.7568233618093956  0.5753862248163557  0.5000000000000000
```

```
  0.7568744457180685  0.6674159599941447  0.5000000000000000
  0.6636424071576812  0.7606364770670879  0.5000000000000000
  0.5715558320596585  0.7606630990069903  0.5000000000000000
  0.5715242213540108  0.6674327733683398  0.5000000000000000
  0.6636000714206318  0.5753859305270055  0.5000000000000000
  0.3317260461979927  0.1189004275888408  0.5000000000000000
  0.5453258849982134  0.1188661693908273  0.5000000000000000
  0.5453103954769887  0.3353701810465424  0.5000000000000000
  0.3316903204497379  0.5489221581163690  0.5000000000000000
  0.1152562850212178  0.5489008884336144  0.5000000000000000
  0.1152774716813028  0.3353169712406583  0.5000000000000000
  0.8794727714269968  0.4532695186094813  0.5000000000000000
  0.8795454974597803  0.6667981476627745  0.5000000000000000
  0.6630890261988966  0.8832837950901862  0.5000000000000000
  0.4494830271996690  0.8833050897355506  0.5000000000000000
  0.4494663317811103  0.6668142835429904  0.5000000000000000
  0.6630877588033925  0.4532678359677433  0.5000000000000000
  0.4974065996314536  0.0010860619143571  0.5000000000000000
  0.4973998373865669  0.5010834981451353  0.5000000000000000
  0.9973979673168856  0.5010520634629785  0.5000000000000000
```

**Pd3C12O12**

```
   1.00000000000000
    13.5230577277511905    0.0000000000000000    0.0000000000000000
     6.7615288639263138   11.7113115291076202    0.0000000000000000
     0.0000000000000000    0.0000000000000000   16.0000000000000000
   C    O    Pd
    12    12     3
Direct
  0.3307175618047106  0.2278589907894069  0.5000000000000000
  0.4371192146105049  0.2278880021996105  0.5000000000000000
  0.4371645376086641  0.3344736341537541  0.5000000000000000
  0.3307949799851969  0.4409264021771654  0.5000000000000000
  0.2242484805215176  0.4409231143899349  0.5000000000000000
  0.2242088190888299  0.3344743460561475  0.5000000000000000
  0.7706182273217120  0.5612601900049569  0.5000000000000000
  0.7705631821368311  0.6677338561448426  0.5000000000000000
  0.6640009209794073  0.7743147179045872  0.5000000000000000
  0.5575917825575580  0.7742825334641392  0.5000000000000000
  0.5576525333418162  0.6676441425649642  0.5000000000000000
```

```
  0.6640953333465544  0.5612026047636872  0.5000000000000000
  0.3351784711510618  0.1294872930747180  0.5000000000000000
  0.5310355173022556  0.1295400700720841  0.5000000000000000
  0.5310902937254679  0.3388861245733498  0.5000000000000000
  0.3353131065262641  0.5348127790115199  0.5000000000000000
  0.1258722325152497  0.5348320813276033  0.5000000000000000
  0.1257914014874520  0.3390487246642705  0.5000000000000000
  0.8689889270262343  0.4673468318487377  0.5000000000000000
  0.8688728438333596  0.6633900015117089  0.5000000000000000
  0.6595216692912843  0.8727227206324559  0.5000000000000000
  0.4636666659004354  0.8726212385494471  0.5000000000000000
  0.4637545030799471  0.6631779635697612  0.5000000000000000
  0.6596436288972206  0.4672489021684137  0.5000000000000000
  0.4973469018811076  0.0010901096918599  0.5000000000000000
  0.4974755480350157  0.5010178648092307  0.5000000000000000
  0.9973816715778121  0.5011854321965176  0.5000000000000000
```

**Pd3C12S12**

```
   1.00000000000000
    15.0738667171281104    0.0000000000000000    0.0000000000000000
     7.5369333586205851   13.0543515103289494    0.0000000000000000
     0.0000000000000000    0.0000000000000000   16.0000000000000000
   C    S    Pd
    12    12     3
Direct
  0.3311737902855921  0.2390254113590311  0.5000000000000000
  0.4255515658533899  0.2390675959140038  0.5000000000000000
  0.4255404355917953  0.3348931185380917  0.5000000000000000
  0.3311782698144738  0.4293177264204702  0.5000000000000000
  0.2354248819190400  0.4292833078745522  0.5000000000000000
  0.2354312906045568  0.3348460768847903  0.5000000000000000
  0.7594509259534759  0.5729205508709466  0.5000000000000000
  0.7594383000325881  0.6673003080511410  0.5000000000000000
  0.6636244649689036  0.7630600230644085  0.5000000000000000
  0.5691928521293680  0.7630859936491359  0.5000000000000000
  0.5692083497191089  0.6672845398400205  0.5000000000000000
  0.6636641863101274  0.5729055738716656  0.5000000000000000
  0.3290761327796823  0.1268923203355337  0.5000000000000000
  0.5397464632507436  0.1269051294143842  0.5000000000000000
  0.5397391604660462  0.3328770229610131  0.5000000000000000
```

```
  0.3291799652148505  0.5435155824152577  0.5000000000000000
  0.1232427345502298  0.5434922161282167  0.5000000000000000
  0.1232096754001147  0.3328872053383849  0.5000000000000000
  0.8716108466764592  0.4586855037859063  0.5000000000000000
  0.8716188547812166  0.6693466878583773  0.5000000000000000
  0.6655646835784971  0.8753006815685239  0.5000000000000000
  0.4550008787102212  0.8752382471476863  0.5000000000000000
  0.4550192731190306  0.6692864418240951  0.5000000000000000
  0.6656586732030610  0.4586937084254321  0.5000000000000000
  0.4973295254417570  0.0010899869339276  0.5000000000000000
  0.4974164787256705  0.5010780929025830  0.5000000000000000
  0.9974162964535421  0.5011116189373457  0.5000000000000000
```

**Pd3C12Se12**

```
   1.00000000000000
     15.7537372951821109    0.0000000000000000    0.0000000000000000
      7.8768686475910750   13.6431367021991896    0.0000000000000000
      0.0000000000000000    0.0000000000000000   16.0000000000000000
   C    Se   Pd
    12    12     3
Direct
  0.3311559238954089  0.2436772885546503  0.5000000000000000
  0.4209508277745186  0.2436393247977051  0.5000000000000000
  0.4209816336321097  0.3348452908238926  0.5000000000000000
  0.3312143380676886  0.4246084319016390  0.5000000000000000
  0.2399688865836040  0.4246542348891111  0.5000000000000000
  0.2399402863260391  0.3348981168359231  0.5000000000000000
  0.7547842892612380  0.5775707255535494  0.5000000000000000
  0.7547964404333030  0.6673355518786792  0.5000000000000000
  0.6635515643136571  0.7585445013109862  0.5000000000000000
  0.5737699521036177  0.7585401485199696  0.5000000000000000
  0.5737653100868343  0.6673201202107322  0.5000000000000000
  0.6635576806679921  0.5775774339278001  0.5000000000000000
  0.3275533708793574  0.1263953414048586  0.5000000000000000
  0.5418873560982860  0.1263983511640845  0.5000000000000000
  0.5419275822667018  0.3311583810142693  0.5000000000000000
  0.3276641546774712  0.5454768741773250  0.5000000000000000
  0.1227712534270040  0.5456016611682983  0.5000000000000000
  0.1226825614324412  0.3312638713527036  0.5000000000000000
  0.8720582760055677  0.4566595293966529  0.5000000000000000
```

```
  0.8720616301038575  0.6709887238200238  0.5000000000000000
  0.6672276939464652  0.8757738906256520  0.5000000000000000
  0.4528873159502567  0.8757864309591739  0.5000000000000000
  0.4529231412955923  0.6708541153335048  0.5000000000000000
  0.6673079968400870  0.4566183013997005  0.5000000000000000
  0.4974052589926501  0.0010868334630700  0.5000000000000000
  0.4975104431076360  0.5009777104458522  0.5000000000000000
  0.9974037873642061  0.5011394873851955  0.5000000000000000
```

**Rh3C12O12**

```
   1.00000000000000
    13.5992963743756796    0.0000000000000000    0.0000000000000000
     6.7996481872388648   11.7773361338347904    0.0000000000000000
     0.0000000000000000    0.0000000000000000   16.0000000000000000
   C    O    Rh
    12    12     3
Direct
  0.3304182251569472  0.2284333855385441  0.5000000000000000
  0.4370384450672447  0.2284190427751938  0.5000000000000000
  0.4370754804007078  0.3340798000092349  0.5000000000000000
  0.3304725621496090  0.4406911566910381  0.5000000000000000
  0.2248048485432577  0.4407029523308594  0.5000000000000000
  0.2247770575674721  0.3341055609699808  0.5000000000000000
  0.7700586885209830  0.5614393237916744  0.5000000000000000
  0.7700284264117059  0.6680699855765724  0.5000000000000000
  0.6643462888935829  0.7737675613066486  0.5000000000000000
  0.5577348389489245  0.7737765348840995  0.5000000000000000
  0.5577834243007018  0.6680660786143804  0.5000000000000000
  0.6643930430466511  0.5614468882120676  0.5000000000000000
  0.3374265909697556  0.1299476564146005  0.5000000000000000
  0.5284585792908771  0.1298981087675486  0.5000000000000000
  0.5285105641024117  0.3411532122129373  0.5000000000000000
  0.3375127103262372  0.5321564911666030  0.5000000000000000
  0.1262678753703810  0.5321335898038910  0.5000000000000000
  0.1262335080924260  0.3412163488863911  0.5000000000000000
  0.8685872708824576  0.4700248381425922  0.5000000000000000
  0.8685557481711044  0.6609654743906077  0.5000000000000000
  0.6571981509375462  0.8723309781687263  0.5000000000000000
  0.4662810329066858  0.8722853707630520  0.5000000000000000
  0.4663527021921183  0.6609583163768278  0.5000000000000000
```

```
  0.6572348310080045  0.4700432093851106  0.5000000000000000
  0.4973509030982868  0.0011041928964559  0.5000000000000000
  0.4974193316780386  0.5010751065605703  0.5000000000000000
  0.9974376414970081  0.5010797592801310  0.5000000000000000
```

**Rh3C12S12**

```
1.00000000000000
    15.0919276840278709    0.0000000000000000    0.0000000000000000
     7.5459638420139372   13.0699927664919908    0.0000000000000000
     0.0000000000000000    0.0000000000000000   16.0000000000000000
   C    S    Rh
    12    12     3
Direct
  0.3307948902906124  0.2394718379819065  0.5000000000000000
  0.4255108210679014  0.2395028036113193  0.5000000000000000
  0.4255332662128244  0.3345187970116456  0.5000000000000000
  0.3308420810791978  0.4292676393429609  0.5000000000000000
  0.2358368723129448  0.4292684166452605  0.5000000000000000
  0.2357988542234701  0.3345335083965503  0.5000000000000000
  0.7590003784671330  0.5729356090186443  0.5000000000000000
  0.7589822703044424  0.6676696754427203  0.5000000000000000
  0.6639513917138837  0.7626898279557540  0.5000000000000000
  0.5691854778396988  0.7627152717176061  0.5000000000000000
  0.5692059128241127  0.6676552620408294  0.5000000000000000
  0.6639917002595432  0.5729312355372019  0.5000000000000000
  0.3314656888518877  0.1256532585139861  0.5000000000000000
  0.5386963386689529  0.1256874348964010  0.5000000000000000
  0.5387544257397018  0.3350857502020639  0.5000000000000000
  0.3315523211720545  0.5424165744352063  0.5000000000000000
  0.1219970617683461  0.5424282264956515  0.5000000000000000
  0.1219609070795086  0.3352340834147854  0.5000000000000000
  0.8728324097158398  0.4597802042906590  0.5000000000000000
  0.8728012716002610  0.6670150239145398  0.5000000000000000
  0.6633120113388449  0.8765199198752690  0.5000000000000000
  0.4560214746271700  0.8765368150025604  0.5000000000000000
  0.4560700489550489  0.6669325564768203  0.5000000000000000
  0.6633895730254665  0.4597037875588557  0.5000000000000000
  0.4973873829812829  0.0010943191728668  0.5000000000000000
  0.4974820916113130  0.5010007572973835  0.5000000000000000
  0.9974018457996828  0.5011223276668202  0.5000000000000000
```

**Rh3C12Se12**

```
1.00000000000000
    15.7683920765507999    0.0000000000000000    0.0000000000000000
     7.8841960382754070   13.6558281151514205    0.0000000000000000
     0.0000000000000000    0.0000000000000000   16.0000000000000000
   C     Se    Rh
    12    12     3
Direct
  0.3309492585867488  0.2443676345596728  0.5000000000000000
  0.4205199808654925  0.2443264111572603  0.5000000000000000
  0.4205470037839163  0.3347225372733220  0.5000000000000000
  0.3309883264188045  0.4242815390132151  0.5000000000000000
  0.2405584154489944  0.4243053899158440  0.5000000000000000
  0.2405299748219321  0.3347841811381755  0.5000000000000000
  0.7542159376371345  0.5778768268291614  0.5000000000000000
  0.7542254180140091  0.6674264762154465  0.5000000000000000
  0.6638016728003251  0.7578273558743973  0.5000000000000000
  0.5742422638331632  0.7578410677648009  0.5000000000000000
  0.5742233006047428  0.6674522263576819  0.5000000000000000
  0.6637844089043554  0.5778982789017064  0.5000000000000000
  0.3300112421915813  0.1246508331170588  0.5000000000000000
  0.5412167199714943  0.1246497603634680  0.5000000000000000
  0.5412501893869432  0.3336802090736626  0.5000000000000000
  0.3300362236011978  0.5449381806513003  0.5000000000000000
  0.1209283541377602  0.5449977464786002  0.5000000000000000
  0.1209069666164027  0.3337346694141132  0.5000000000000000
  0.8739075519690331  0.4572262994808582  0.5000000000000000
  0.8738731114231726  0.6684703875076607  0.5000000000000000
  0.6647941424167101  0.8775127300919792  0.5000000000000000
  0.4535938319265611  0.8775345270864889  0.5000000000000000
  0.4535737722355293  0.6684155095188089  0.5000000000000000
  0.6648149857199499  0.4572076947650885  0.5000000000000000
  0.4974157212954537  0.0010842337109622  0.5000000000000000
  0.4974408363852021  0.5010407772746390  0.5000000000000000
  0.9974091585345235  0.5011174403808596  0.5000000000000000
```

## 2. AIMD results for the thermal stability

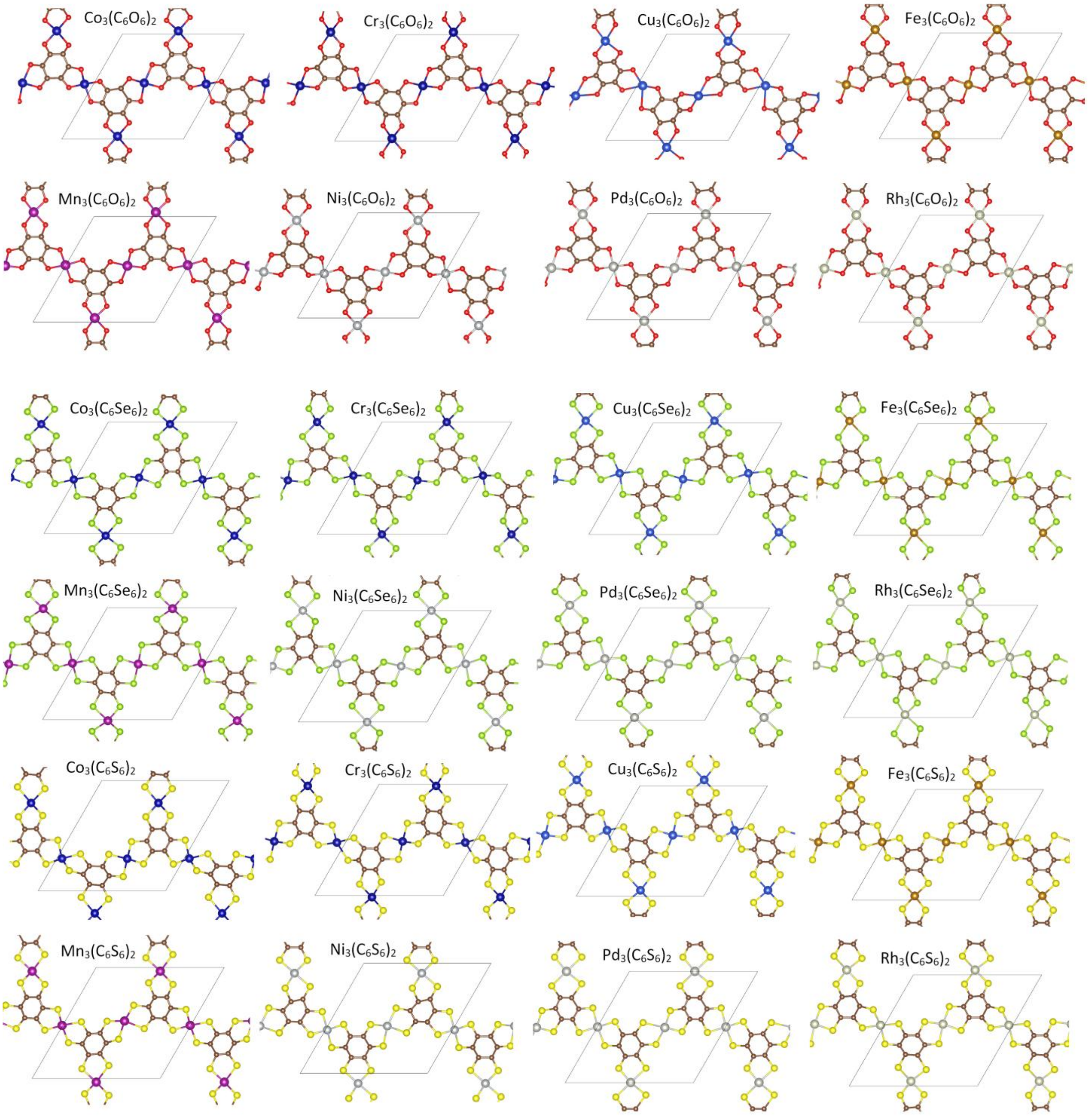

Fig. S1, Top views of MOF monolayers at 1500 K after the AIMD simulations for 15 ps.